\documentclass[11pt]{article}
\usepackage{jheppub}

\usepackage{epsfig}
\usepackage{amssymb}
\usepackage{amsfonts}
\usepackage{amsbsy}
\usepackage{amsmath}
\usepackage{dsfont}
\usepackage{bbm}
\usepackage{upgreek}
\usepackage{amssymb,amscd}
\usepackage{graphicx}
\usepackage{mathrsfs}
\usepackage{amsmath,amsthm}
\usepackage{slashed}
\usepackage{dsfont}
\usepackage{tikz}
\usepackage[utf8]{inputenc}
\makeatletter
\DeclareFontFamily{OMX}{MnSymbolE}{}
\DeclareSymbolFont{MnLargeSymbols}{OMX}{MnSymbolE}{m}{n}
\SetSymbolFont{MnLargeSymbols}{bold}{OMX}{MnSymbolE}{b}{n}
\DeclareFontShape{OMX}{MnSymbolE}{m}{n}{
    <-6>  MnSymbolE5
   <6-7>  MnSymbolE6
   <7-8>  MnSymbolE7
   <8-9>  MnSymbolE8
   <9-10> MnSymbolE9
  <10-12> MnSymbolE10
  <12->   MnSymbolE12
}{}
\DeclareFontShape{OMX}{MnSymbolE}{b}{n}{
    <-6>  MnSymbolE-Bold5
   <6-7>  MnSymbolE-Bold6
   <7-8>  MnSymbolE-Bold7
   <8-9>  MnSymbolE-Bold8
   <9-10> MnSymbolE-Bold9
  <10-12> MnSymbolE-Bold10
  <12->   MnSymbolE-Bold12
}{}

\let\llangle\@undefined
\let\rrangle\@undefined
\DeclareMathDelimiter{\llangle}{\mathopen}%
                     {MnLargeSymbols}{'164}{MnLargeSymbols}{'164}
\DeclareMathDelimiter{\rrangle}{\mathclose}%
                     {MnLargeSymbols}{'171}{MnLargeSymbols}{'171}
\makeatother

\def\be{ \begin{equation} }
\def\ee{ \end{equation}}

\newcommand{\eq}[1]{\begin{align}\begin{split}#1\end{split}\end{align}}

\def\cot{{\rm cot}}

\def\log{{\rm log}}

\def\half{\frac{1}{2}}

\def\ihalf{\frac{i}{2}}

\def\Nf{{N_f}}

\def\one{{\hbox{ 1\kern-.8mm l}}}

\def\vS{{\vec{S}}}
\def\vL{{\vec{L}}}

\def\CD {{\cal D}}

\def\CH {{\cal H}}

\def\CK {{\cal K}}
\def\CL {{\cal L}}

\def\CH {{\cal H}}

\def\CU {{\cal U}}

\def\IR{{\mathbb{R}}}

\def\IZ{{\mathbb{Z}}}

\def\rmk#1{\bigskip\noindent{\bf Remark} }
\def\cnj#1{\bigskip\noindent{\bf Conjecture:} }

\def\tildeH{{\widetilde{H}}}

\DeclareMathAlphabet{\mathpzc}{OT1}{pzc}{m}{it}

\def\Tr{ \, \textrm{Tr} \, }

\def\csch{ \, \textrm{csch} \, }

\title{A New Solution to the Callan Rubakov Effect}

\author{T.~Daniel Brennan}

\affiliation{Department of Physics, University of California San Diego,\\
 \textit{9500 Gilman Drive, La Jolla CA 92093-0319, USA}}

\emailAdd{tbrennan@ucsd.edu}

\abstract{ In this paper we study the scattering of massive fermions off of smooth, spherically symmetric monopoles in $4d$ $SU(2)$ gauge theory.  
We propose a complete physical picture of the monopole-fermion interaction which encompasses all angular momentum modes.  
We show that as an in-going fermion scatters off  a monopole, it excites trapped $W$-bosons in the monopole core by a version of the Witten effect so that the monopole can accrue charge and transform into a dyon at parametrically low energies. 
The imparted electric charge is then protected from decay by an emergent $\mathbb{Z}_N$ generalized global symmetry, creating a stable dyon.  
At sufficiently low energies, the scattered fermion can be trapped by the dyon's electrostatic potential, forming a bound state, which can decay into spherically symmetric fermion modes subject to the preserved $\mathbb{Z}_N$ global symmetry. 
We propose that monopole-fermion scattering can be described in this way without needing to add ``new'' states to the Hilbert space, thereby eliminating a long standing confusion in the Callan Rubakov effect. 
}

\begin{document}

\maketitle
 
\section{Introduction}
 
Smooth monopoles are non-perturbative excitations of non-abelian gauge theories that play an important role in both particle and formal high energy theory. 
Since the discovery that monopoles occur naturally in grand unified theories (GUT theories) \cite{tHooft:1974kcl,Polyakov:1974ek}, it has been realized that there are many ways in which monopoles can lead to  measurable effects. One such way is known as the Callan Rubakov effect \cite{Callan:1982au,Callan:1982ah,Callan:1982ac,Rubakov:1982fp,Rubakov:1981tf,Rubakov:1988aq,Preskill:1984gd,Yamagishi:1982wp,Grossman:1983yf,Yamagishi:1983ua,Yamagishi:1984zu,Panagopoulos:1984ws,Besson:1981ch,Dixon:1984xh,Isler:1987xn,Isler:1986ii,Sen:1984qe,Sen:1984kf,Balachandran:1983sw,Nair:1983ps,Nair:1983hc,Smith:2019jnh,Smith:2020nuf,Nelson:1983bu,Affleck:1993np,Hortacsu:1984gx,Hortacsu:1984bx}. 

The Callan Rubakov effect describes the interaction between (massless) fermions and a smooth, spherically symmetric monopole in $4d$ gauge theory. The physically interesting features of this interaction come from the fact that the fermions  
probe the UV physics inside the monopole core. In the real universe, the Callan Rubakov effect is thought to appear in the interaction of Standard Model fermions with GUT monopoles, which gives rise to baryon number violating interactions that lead to proton decay.  

Recently the Callan Rubakov effect has been revisited with renewed interest due to a long standing confusion of the monopole-fermion interaction \cite{Kitano:2021pwt,Csaki:2022qtz,Hamada:2022eiv,Brennan:2021ewu,vanBeest:2023dbu,Csaki:2020inw,Csaki:2022tvb,Csaki:2022qtz}. The feature of interest is that an incoming particle appears to produce a collection of ``fractional'' out-going particles. This effect goes all the way back to the original analysis by Callan \cite{Callan:1982au} and has lead to a myriad of attempts to explain the physics behind this computation.

Much of the original analysis is based on the constraints imposed by the global symmetries that are preserved by the scattering process. As with all physical processes, the scattering preserves gauge ``symmetry.'' Additionally, the scattering process preserves angular momentum, which is especially simple for the case of spherically symmetric monopoles as we consider here. These constraints are also often supplemented with the expectation that the asymptotic states should be matched by the free field quantization since the energy density in the electromagnetic field falls off like $1/r^4$ so that fields are well approximated by free fields far away from the magnetic monopole. 

There are roughly three different classes of interpretations of the fractional asymptotic states that arise in the Callan Rubakov effect:
\begin{itemize}
\item The fractional states are interpreted as a non-trivial density matrix \cite{Callan:1983tm}. This proposal has the problem that it can lead to gauge charges that are only conserved up to classical probability.

\item There are asymptotic states in the monopole Hilbert space that differ from that of free field quantization. These asymptotic states have been attributed to different physical effects such as ``vacuum polarization/soft fermionic radiation'' \cite{Polchinski:1984uw,Brennan:2021ewu}, ``pancakes'' \cite{Kitano:2021pwt}, attachment of topological lines \cite{vanBeest:2023dbu}, or simply that the in- and out-going Hilbert spaces are in different super-selection sectors parametrized by a $U(1)$-valued parameter \cite{Hamada:2022eiv}. These proposals claim that the asymptotic structure of QED differs from the free field quantization in the presence of a monopole.\footnote{
A point that contributes to the confusion of the analysis of the second option is that all symmetries acting on the fermions that are not violated by an Adler-Bell-Jackiw (ABJ) anomaly \cite{Adler:1969gk,Bell:1969ts} (which can be perturbatively activated in a monopole background) are conserved and that all asymptotic states have integer charges with respect to these symmetries. This makes it particularly tempting to conclude that the fractional fermion states do in fact exist in nature.  }  

\item The ``fractional fermion states'' are purified by higher angular momentum states since the total angular momentum of a multi-particle state can be constructed from single-particle states with higher angular momentum \cite{Csaki:2022qtz}. However, it is unclear what interaction would give rise to this entangling effect given that the fermion-monopole interaction is spherically symmetric. 
\end{itemize}

\subsection{Outline and Summary of Results}

In this paper we will present an alternative proposal to describe the physics of the Callan Rubakov effect. 
Here we will argue that there is an additional physical mechanism that allows us to describe the scattering process in terms of the asymptotic states from free field quantization while preserving all global symmetries.  

To perform our analysis, we will take a slightly different approach  
and consider \textit{massive fermions} interacting with spherically symmetric monopoles. The reason is two-fold. First, massless charged fields are subtle and much of our standard intuition from particle physics does not apply. For example, classical massless charged particles do not have bound states in a background electric charge. Perhaps more importantly, massless QED is not IR finite\footnote{Here we differentiate between massive QED which is IR finite in the sense that Fadeev-Kulish asymptotic scattering states \cite{Kulish:1970ut} can be well defined for massive QED  but  are singular for massless QED \cite{Prabhu:2022zcr}.
} and so it is possible that the confusion with understanding the Callan Rubakov effect may come from unregulated IR divergences. 
The second reason we consider massive fermions is that they are  relevant for particle phenomenology. 

{An important issue that arises in using massive fermions is that the Callan-Rubakov effect is only required by symmetry for massless fermions. For massless fermions, there is a large number of chiral symmetries (one of which carries an ABJ anomaly) which are broken when introducing a fermion mass. These enhanced symmetries forbid symmetry preserving purely fermionic boundary conditions, but do admit symmetry preserving boundary conditions if one introduces fractional fermionic states. On the other hand, the reduced symmetry of massive fermions do admit symmetry preserving boundary conditions and therefore generally does not require fractional fermion states. In other words, symmetry requires the Callan Rubakov effect for massless fermions, but not for massive fermions. However, in a UV complete $4d$ gauge theory as we study here, the monopole-fermion interaction which gives rise to the IR boundary condition will be independent of whether or not the fermion is massive. This is consistent with effective field theory when we scatter at sufficiently high energies comparable to the fermion masses $m_W\gg E \gg m_\psi$. In this limit, we expect that the monopole-fermion scattering is insensitive to the fermion mass and should behave as if they were massless fermions, thereby reproducing some of the salient features of the Callan-Rubakov effect for massless fermions. We would again like to emphasize that we will not solve the ``massless Callan-Rubakov effect'' but we will find that studying the ``massive Callan-Rubakov effect'' will give insights into the massless case. 
}

In this paper we will restrict our analysis to $SU(2)$ QCD with $N_f{=2n_f}$ fundamental Weyl fermions with mass $m_\psi$. If we couple this theory to an adjoint scalar field with potential, then RG flows along which the scalar condenses above the confinement scale will lead to a $U(1)$ gauge theory with charged fermions at energies below the non-abelian gauge boson mass (i.e. $W$-boson mass): $E\ll m_W$. Here we assume $m_W>>m_\psi$. In this theory, there is a single spherically symmetric, smooth monopole when we spontaneously break $SU(2)\to U(1)$ which flows to the minimal monopole line in the IR $U(1)$ gauge theory. This monopole contains a trapped $W$-boson inside of its core which in the low energy effective theory gives rise to a charged $U(1)$-valued scalar field (i.e. the phase mode of the trapped $W$-boson) on the monopole world volume called the ``dyon degree of freedom'' which is governed by a $1d$ quantum mechanics of a particle of mass $m_W$ on a ring. 

We propose that the scattering of charged fermions off of the monopole occurs as follows. 
An incident  charged fermion will scatter off of the monopole by interacting with the trapped $W$-boson via the UV tree level interaction.  
This excites the dyon degree of freedom and  deposits charge on the monopole, turning it into a dyon of charge 2. Naively, the dyon degree of freedom is heavy and requires energy of the order of the GUT scale -- $E_{GUT}\sim m_W\gg m_\psi$ -- to be excited. For this reason, it has often been assumed that this charge must be radiated away. 

However, our proposal is that the fermion can deposit charge on the monopole at energies that are parametrically lower than $m_W$. Physically, this occurs by a version of the Witten effect. In terms of the quantum mechanics governing the dyon degree of freedom scattering the fermions off of the monopole induces a non-trivial $1d$ $\theta$-angle. The induced $\theta$-angle leads to a shift of the energy levels  so that the lowest energy state sources a bulk electric field. 

This reflects the fact that fermion scattering effectively induces a shift in the $4d$ $\theta$-angle on the monopole. The shift in the $\theta$-angle is induced by the phase mode of the fermions which  acts as an axion field \cite{Peccei:1977ur,Peccei:1977hh,Kim:1979if,Shifman:1979if} that couples to the gauge field according to the transformation properties of the associated fermion. 
 In fermion scattering, the axion winds, shifting the fermion vacuum which induces a vacuum charge density of radius $R_c\sim 1/m_\psi$: it contributes $E\sim m_\psi$ to the mass of the dyon instead of $E\sim m_W$. This is the reason why the monopole can be transformed into a dyon by parametrically low energies. 

The electric charge that is deposited on the monopole is in fact protected from decay by an emergent $\IZ_{N_f}$ generalized global symmetry which we will denote as $\IZ_\Nf^{(0,1)}$. This symmetry reflects the fact that there is no way of distributing a non-zero charge amongst the symmetrically coupled $N_f$ fermions unless the charge is a multiple of $N_f$ without breaking the  $SU(N_f)$ global  symmetry. This symmetry prevents the creation of asymptotic states that have fractional fermion number. 

Since each scattering deposits 2-units of electric charge, our proposal leads to charge accumulation on the monopole when we scatter less than $\frac{\Nf}{2}$ fermions off of the monopole.  However, when we scatter $\frac{\Nf}{2}$ fermions off of the monopole, the $\IZ_{N_f}^{(0,1)}$ symmetry does not protect the dyon from decay and the charge is indeed radiated away. This reproduces the standard Callan Rubakov effect for asymptotic states with integer fermion number. This can also be seen by integrating out the dyon degree of freedom as in \cite{Fan:2021ntg} which leads to an `t Hooft vertex that is localized on the monopole world-volume. This interaction relates $\frac{\Nf}{2}$ in-coming fermion modes to $\frac{\Nf}{2}$ out-going fermion modes. 
This argument is in the spirit of the original analysis of Rubakov \cite{Rubakov:1982fp}. 

If the incident fermion does not have sufficient energy, the scattered fermion can be trapped by the electrostatic potential of the dyon, forming a bound state of radius $R_0$ which scales parametrically with the fermion mass as  $R_0\sim 1/{m_\psi}$ -- similar to the hydrogen atom \cite{Zhang:1988ab,Ravendranadhan:1989vp,Tang:1982fc}. 

We would like to make a couple important comments about our proposal: 
\begin{itemize}

\item Even though the higher angular momentum modes have no (plane wave) normalizable states with energy $E=m_\psi$ \cite{Brennan:2021ucy}, the $j>0$ modes do not have an energy gap. This means that there are higher angular momentum modes in the low energy effective theory that can couple to the dyon degree of freedom.

Furthermore, the higher angular momentum modes must couple to the dyon degree of freedom. This can be seen from the fact that the independent modes in the non-abelian monopole background have a net asymptotic charge flux. This implies that in the IR effective theory an in-going scattering state is reflected off of the monopole core into an out-going scattering state of opposite electric charge so that there is a net electric charge deposited on the monopole core, which requires interacting with the dyon degree of freedom.  

The fact that the higher angular momentum modes couple to the dyon degree of freedom should be anticipated from the results of \cite{Brennan:2021ewu} which studies the Callan Rubakov proposal in general $SU(N)$ gauge theories with fermions in generic representations and higher charge spherically symmetric monopoles. When considering more general spherically symmetric monopoles and fermion representations, the Callan Rubakov  effect crucially relies on the existence of higher orbital angular momentum modes (although not higher total angular momentum modes) which experience the typical orbital angular momentum barrier. It is usually argued that this barrier prevents the modes from reaching the monopole core, however the results of \cite{Brennan:2021ewu} show that this would lead to a violation of gauge symmetry in the Callan Rubakov effect.

\item We would like to further emphasize that our proposal preserves all conserved symmetries and completely matches with the proposal of Callan and Rubakov for scattering processes with asymptotic states that have integer fermion number. 

In the case of the $SU(5)\to\frac{SU(3)\times SU(2)\times U(1)}{\IZ_6}$ GUT theory with a \textit{single generation}, the minimal monopole can trap a single anti-proton:
\eq{
\bar{p}+M~\longrightarrow (D+2u+e^-)~,
}
at low enough scattering energies, 
but does not trap pairs of anti-protons: 
\eq{
2\bar{p}+M~\longrightarrow~ 3e^-+p+M~.
}
These processes both lead to baryon and lepton number violation. Note that in the first process, the dyon  carries $U(1)_{B-L}$ charge due to the fact that the generator of the $U(1)_{B-L}$ in the $SU(5)$ theory is given by 
\eq{
Q_{B-L}=Q_Y+Q_D
}
where $Q_Y$ is the generator of hypercharge and $Q_D$ acts on the $SU(5)$ fermion multiplets with charges $Q_D[\mathbf{\bar{5}}]=-3$ and $Q_D[\mathbf{10}]=1$. Since the monopole does not carry charge with respect to $Q_D$, but does with respect to $Q_Y$, the dyon can carry charge with respect to $Q_{B-L}$. When enough charge is accumulated, the $U(1)_{B-L}$ charge is radiated away, and thus $U(1)_{B-L}$ is conserved as in the original Callan Rubakov effect.

\item For technical reasons, we will restrict to the limit where $N_f<<\frac{1}{g^2}$ where $g$ is the gauge coupling which is parametrically weak. Here, our assumptions allow us to ignore loop corrections and prevents the screening of IR electric charges as discussed in \cite{Aharony:2022ntz}. Although we will comment on these effects, we will leave a detailed analysis of these effects for future investigation.

\end{itemize}

\bigskip
The outline of the paper is as follows. In Section \ref{sec:Monopoles}, we will begin by giving some background on the spherically symmetric monopoles in $4d$ $SU(2)$ QCD theory with $N_f$ Weyl fermions transforming in the fundamental representation of the gauge group. There we will describe the mode expansion of the fermion fields in the monopole background and demonstrate the lack of an energy gap for the higher angular momentum modes. 
Then in Section \ref{sec:symmetries} we will study the symmetries of the theory in both the UV and IR and argue for our proposal from analyzing the constraints from global symmetries. Then in Section \ref{sec:2D} we will study the scattering of s-wave  fermions off of the monopole which can be described by an effective $2d$ theory. We will derive our new mechanism and demonstrate how monopoles can capture fermions. 
Next, in Section \ref{sec:GeneralScattering} we will 
generalize the $2d$ s-wave analysis to all angular momentum modes and describe 
general monopole-fermion scattering.  
Finally, we conclude by discussing the general application to Standard Model phenomenology and other future directions.

\section{Monopoles and Fermions in $SU(2)$ Yang-Mills}
\label{sec:Monopoles}

The setting for our paper is $4d$ $SU(2)$ gauge theory coupled to an adjoint scalar field $\Phi$ and $N_f$ Weyl fermions $\psi$ that transform in the  fundamental representation. Here we include a potential for $\Phi$ so that it condenses above the confinement scale. 
This spontaneously breaks the gauge field $SU(2)\to U(1)$ and the theory flows to QED. 

When the gauge symmetry is spontaneously broken $SU(2)\to U(1)$, the theory has smooth dynamical monopoles of mass $m_{\rm mono}\sim m_W/g^2$ where $m_W$ is the mass of the $W$-boson (i.e. massive gauge boson) and $g$ is the gauge coupling.
Here we will be focusing only on spherically symmetric monopoles. 

 For $SU(2)$ gauge theory, there is a unique spherically symmetric monopole. Spherically symmetric monopoles are those which are rotationally invariant up to a gauge transformation. In other words, spherically symmetric monopoles are invariant under the modified generator of angular momentum:
\eq{\label{Jnonabelian}
\vec{J}=\vec{L}+\vec{S}+\vec{T}~,
}
where $\vec{T}$ are the generators of an $SU(2)\subset G_{\rm gauge}$ and $\vL,\vS$ are the standard generators of orbital- and spin-angular momentum. The spherically symmetric  monopole is a Yang-Mills-Higgs field configuration, which in the Parasad-Sommerfield limit\footnote{This is the limit in which the Higgs potential is set to zero while fixing the Higgs vev. }  satisfies
\eq{
B_\mu=D_\mu\Phi~. 
}
The explicit solution of this equation can be written as 
\eq{\label{Connection}
&A_{\rm mono}=T_3 A_{\rm Dirac}+\ihalf T_+\, M(r)\, e^{ - i \phi}(d\theta-i \sin\theta d\phi)-\ihalf T_-\, M(r)\, e^{ i \phi}(d\theta+i \sin\theta d\phi)~,\\
&\Phi= T_3 h(r)~,
}
where {$T_\pm=T_1\pm i T_2$  and the above functions are given }
\eq{\label{BPSprofile}
M(r)=m_W r\, \csch(m_W r)\quad, \quad h(r)=m_W \coth(m_W r)-\frac{1}{r}~,}
{where} $A_{\rm Dirac}$ is the Dirac monopole connection.\footnote{Explicitly $A_{\rm Dirac}=\half(\sigma-\cos(\theta))d\phi$ where $\sigma=\pm 1$ on the northern/southern hemisphere. For simplicity in this paper  we will work in the northern hemisphere coordinate patch.} For simplicity, we will generally restrict ourself to the Prasad-Sommerfield limit because 1.) {we expect} the fermion-BPS monopole interaction captures the physics of the general case and 2.) we have explicit forms for the monopole connection in this limit. \footnote{{The effect of going away from the BPS limit is that the Higgs potential will ``squeeze'' the monopole core so that the monopole radius is smaller than the BPS monopole. However, this difference should be unnoticed by the IR observer to leading order.} 
}

 Here we can see from the explicit form of $h(r)$ that although the Higgs vev gives a mass to the $W_\pm$-bosons, the mass goes to zero inside the core of the monopole. 
  This localizes a charged degree of freedom to the monopole core which is called the 
``dyon degree of freedom'' that is described by a $1d$ charged periodic scalar field $\varphi(t)$. 
In fact, as discussed in \cite{Brennan:2021ewu}, all other non-abelian degrees of freedom of the gauge field (that is all other degrees of freedom except the IR photon) acquire a mass of order $m_W$ due to a combination of the profile of $M(r)$ and $h(r)$ in \eqref{BPSprofile}. 

Explicitly, the dyon degree of freedom can be expressed in terms of the (dynamical) phase of the monopole field configuration:
\eq{\label{Adyon}
A_{\rm mono}=T_3 A_{\rm Dirac}\pm \ihalf T_\pm\, M(r)\, e^{ \pm i(\varphi- \phi)}(d\theta\mp i \sin\theta d\phi)~, 
}
where the signs are all correlated and summed over. 
Because of the form of the connection, we can interpret $\varphi(t)$ as the dynamical phase of a trapped $W$-boson and the profile of $M(r)$ exponentially localizes the dyon degree of freedom to the monopole core. 
Given this parametrization of $\varphi(t)$, it is clear that it transforms as a Goldstone mode under the bulk $U(1)$ gauge symmetry due to the fact that the IR gauge transformations $e^{ i \alpha T_3}$ shift $\varphi\to \varphi+2\alpha$. 

Note that we could also infer the existence of the dyon degree of freedom from consistency of the IR theory. The reason is that in the IR, the monopole line is a defect operator which is defined by excising an infinitesimal 3-ball and imposing boundary conditions for the magnetic field there. However, one cannot impose simultaneous boundary conditions on both the electric and magnetic field because they are not   commuting operators. Therefore, when we impose boundary conditions that fix the magnetic field, we must also allow for a electric degree of freedom on the boundary; i.e.  the dyon degree of freedom. Said differently, magnetic boundary conditions are Dirichlet on the gauge field $A$ which can only be made gauge invariant by introducing a charged $U(1)$ scalar degree of freedom on the boundary.

If we plug in the background gauge configuration \eqref{Adyon} into the action, the dynamics of the dyon degree of freedom will be effectively described a $1d$ quantum mechanics of a charged particle on a ring:
 \eq{\label{OGQM}
S_{1d}=\int  
 \frac{1}{2m_W}(\dot\varphi-2a_0)^2dt~,
}
where $\varphi\sim \varphi+2\pi$ and $a_0=A_0\big{|}_{r=0}$. 
The associated Hamiltonian is given in terms of the momentum operator $p=\frac{1}{m_W}(\dot\varphi-2a_0)$: 
\eq{
H=\frac{m_W}{2}p^2+2a_0 p~.
}
In the presence of a trivial electric field, the Hilbert space can be graded by momentum eigen-states
\eq{
\CH={\rm span}\Big\{|n\rangle~,~n\in \IZ~\Big{|}~ \hat{p}|n\rangle=n|n\rangle\Big\}~,
}
which are quantized due to the periodicity of $\varphi$. 
The energy of these states are given by 
\eq{
\hat{H}|n\rangle=\frac{m_W}{2}n^2 |n\rangle~.
}
The coupling between $a_0$ and $p$ 
indicates that the momentum eigenstates source a bulk electric field: the state $|n\rangle$ sources a bulk electric field corresponding to a point particle of charge $2n$.

 It is important for our discussion later that this theory admits a topological $\theta$-term:
\eq{
L=\frac{1}{2m_W}(\dot\varphi-2a_0)^2+ \frac{\theta }{2\pi}(\dot\varphi-2a_0)~. 
}
This deformation leads to a shift in the momentum operator $\hat{p}=\frac{1}{m_W}(\dot \varphi-2a_0)+\frac{\theta}{2\pi}$, without changing the quantization of the states. However, the topological $\theta$-term does shift the energy levels of the system from before
\eq{\label{QMEnring}
E_n=\frac{m_W}{2}\left(n-\frac{\theta}{2\pi}\right)^2~,
}
so that shifting $\theta\to \theta+2\pi$ also shifts the energy levels $E_n\mapsto E_{n+1}$. Thus, if we have an initial $\theta=0$, then shifting $\theta\to \theta+2\pi$ changes the ground state from the $|0\rangle$ state to the $|1\rangle$ state which sources a bulk electric charge.

\subsection{Fermions and the Dirac Monopole}

Now we wish to study the behavior of the fermions in the presence of the spherically symmetric $SU(2)$ monopole. In the IR, this is governed by the interaction of the fermions with the Dirac monopole plus possible boundary conditions/interactions with the dyon degree of freedom. To determine this boundary interaction, we need the fact that the photon and $\varphi(t)$ are the only low energy degrees of freedom of the full non-abelian gauge field. This means that at low energies,  
the  background monopole connection is rigid (up to photon and $\varphi(t)$ fluctuations)  
and therefore 
in the IR, we should expand the low energy fermions in terms of the eigen-spinors of the Dirac operator with the fixed monopole background.\footnote{Said differently, there is a 1-form magnetic symmetry that protects the monopole connection from being screened. It is therefore appropriate to expand the fermion fields in terms of eigen-spinors of the Dirac operator in this fixed background. }
  We can then compute the interaction between the low energy fermions and the dyon degree of freedom by matching the mode expansion of the fermions in the Dirac and non-abelian monopole backgrounds.  
  
In order to facilitate our analysis, we will first consider the IR fermion modes. In the IR, $SU(2)$ QCD flows to QED with $N_f$ Weyl fermions each of charge $+1$ and $-1$. Additionally, at long distances  the smooth, spherically symmetric non-abelian monopole flows to the minimal singular monopole which has the connection 

\eq{A_{\rm Dirac}=\half(\sigma-\cos(\theta))d\phi~,}
and $\sigma=\pm 1$ on the northern/southern hemisphere. As before, we will restrict our analysis to the northern hemisphere coordinate patch -- similar results will hold in the southern hemisphere which is related by a gauge transformation $g=e^{ -i \phi}$ along an equatorial patch.  

It is now our goal to compute the spectrum of the Dirac operator in the singular monopole background above. This monopole background is also spherically symmetric with respect to the modified angular momentum generators \cite{Wu:1976ge,Wu:1977qk,Brennan:2021ucy}
\eq{
\vec{J}=\vec{L}+\vec{S}+\half \hat{r}~.
}
See \cite{Brennan:2021ucy} to see how these angular momentum generators are related to those of the full non-abelian monopole. Since this monopole is spherically symmetric, we can decompose the eigenfunctions of the Dirac operator in terms of angular momentum representations.  We will label these representations by quantum numbers $j,m$:
\eq{
J^2|j,m\rangle=j(j+1)|j,m\rangle\quad, \quad J_3|j,m\rangle=m |j,m\rangle~.
}

Let us first find the spectrum of the massless Dirac operator. Here we will consider a single Weyl fermion $\psi_q$ where $q=\pm1$ denotes the gauge charge.  
 Following from the analysis in \cite{Moore:2014jfa,Brennan:2021ucy}, the Dirac equation with background gauge field \eqref{Connection} can be simplified by performing a frame rotation by 
\eq{
\psi_q=U\hat\psi_q\quad, \quad U =e^{\frac{i \phi}{2}\sigma^0\bar\sigma^3}e^{\frac{i\theta}{2}\sigma^0\bar\sigma^2}~.
}
The resulting Dirac equation is given by   
\eq{
\Bigg[&-i E +\sigma^3\left(\partial_r+\frac{1}{r}\right)+\frac{\CK}{r} 
\Bigg]\hat\psi_q=0~,
}
where 
\eq{\label{ckdef}
\CK&=
\sigma^1\left(\partial_\theta+\half \cot(\theta)\right)+\frac{\sigma^2}{\sin\theta}\Big(\left(\partial_\phi+i \frac{q}{2}\right)-i \cos(\theta)\frac{q}{2}\Big)~,
}
and $q$ is the electric charge of the fermion. 
We can then expand $\psi_q$ into angular momentum eigenstates for $q=\pm1$ as 
\eq{
\hat\psi_{+1}= \left(\begin{array}{c}
f_{j,+} (r)\,e^{i \phi/2} \CD^{(j)}_{m,1}(\theta,\phi)\\
g_{j,+} (r)\,e^{-i \phi/2} \CD^{(j)}_{m,0}(\theta,\phi)
\end{array}\right)\quad, \quad 
\hat\psi_{- 1}= \left(\begin{array}{c}
f_{j,-} (r)\,e^{i \phi/2} \CD^{(j)}_{m,0}(\theta,\phi)\\
g_{j,-} (r)\,e^{-i \phi/2} \CD^{(j)}_{m,-1}(\theta,\phi)
\end{array}\right)~,
}
where   $\CD^{(j)}_{m,q}(\theta,\phi):=\CD^{(j)}_{m,q}(\phi,\theta,\phi)$ are Wigner $D$-functions.\footnote{{
These solutions are written in terms of the Wigner $D$-functions to match with the analysis of \cite{Brennan:2021ucy}. The Wigner $D$-functions are a simple rescaling of the often used Wu-Yang monopole spherical harmonics \cite{Wu:1976ge,Wu:1977qk}:
\eq{
Y_{q,\ell,m}(\theta,\phi)=\sqrt{\frac{2\ell+1}{4\pi}}e^{i(q+m)\phi}d^\ell_{-m,q}(\theta)=\sqrt{\frac{2\ell+1}{4\pi}}\CD^{(\ell)}_{-m,q}(-\phi,\theta,\phi)~.
}
}} Note that when $j=0$, $\CD^{(0)}_{m,\pm1}$ is singular so $f_{0,+} ,g_{0,-} :=0$ and the Dirac equation further simplifies. 

Now, the Dirac equation reduces to a matrix ODE 
\eq{\label{finalDirac}
\left(\sigma^3\left(\partial_r+\frac{1}{r}\right)+\frac{1}{r}\left(\begin{array}{cc}
0&-\lambda\\
\lambda &0
\end{array}\right)-iE\right)\Psi_q=0
}
for a spin-$j$ eigen-spinor and 
\eq{
\Psi_{\pm 1}=\left(\begin{array}{c}
f _{j,\pm}(r)\\
g _{j,\pm}(r)
\end{array}\right)
}
where $\lambda=\sqrt{j(j+1)}$. For $E=0$, the solutions take the simple form
\eq{\label{DiracE0}
\Psi_q=a_1\frac{(m_Wr)^\lambda}{r}\left(\begin{array}{c}
1\\1
\end{array}\right)+a_2\frac{(m_Wr)^{-\lambda}}{r}\left(\begin{array}{c}
1\\-1
\end{array}\right)~, 
}
for $j>0$ and 
\eq{\Psi_q=\frac{1}{r}\begin{cases}\left(\begin{array}{c}
0\\1
\end{array}\right)&q=1\\
\left(\begin{array}{c}
1\\0
\end{array}\right)&q=-1
\end{cases}
}
for $j=0$. 
In order to assess the normalizability of these solutions, we need to use the fact that we are in an IR effective theory with an effective boundary at some finite boundary near $r\to 0$. Thus, the normalizability will be determined by the asymptotic behavior of the solutions at $r\to\infty$ in addition to the boundary conditions at $r\to 0$. For our current discussion, we will only impose normalizability for $r\to \infty$ and leave the discussion of the conditions at $r\to 0$ for the next section. \footnote{Here the correct notion of the normalizable is plane-wave normalizable.} 
Thus, for $E=0$, we will discard the $a_1$ solutions for $j>0$ since they are non-normalizable as $r\to \infty$. 

 For $E>0$, the solutions are formally given by solving the matrix ODE:
\eq{
\Psi_q =P\, e^{-\int \CD_E(r')\,dr'}\Psi_{0,q} \quad, \quad \CD_E(r)=\left(\begin{array}{cc}
-i E+\frac{1}{r}&\frac{\lambda}{r}\\ 
\frac{\lambda}{r}&iE+\frac{1}{r} 
\end{array}\right)~. 
}
To see which of these solutions are normalizable, we again need to consider the $r\to \infty$ limit.  
At $r\to \infty$, the Dirac equation can be written as 
\eq{
\left(\sigma^3\left(\partial_r+\frac{1}{r}\right)-iE\right)\Psi_q =0~,
} 
which has plane wave solutions: 
\eq{\label{asymptoticmassless}
\Psi_q \underset{r\to \infty}{\longrightarrow} \frac{e^{ i E(t+\sigma^3 r)}}{r}\Psi_{0,q} ~. 
}
where $\Psi_{0,{\rm IR}}$ is a constant spinor. 
This means that all solutions are asymptotically plane-wave normalizable for $E>0$. 
 
This means that in addition to the $E=0$ modes, the normalizable, non-negative energy solutions of the Dirac equation are those for which $E>0$ and $\Psi_{0,{\rm IR}}$ is given by 
\eq{\label{constfermIR}
\Psi_{0,q}= c_1\Psi_{0,q}^{(1)}+c_2\Psi_{0,q}^{(2)}=c_1\left(\begin{array}{c}
1\\0
\end{array}\right)+c_2\left(\begin{array}{c}
0\\1
\end{array}\right) 
~,}
for $j>0$ and 
\eq{
\Psi_{0,q}=\begin{cases}
\left(\begin{array}{c}
0\\1
\end{array}\right)&q=1\\
\left(\begin{array}{c}1\\0
\end{array}\right)&q=-1
\end{cases}
} 
for $j=0$ which respectively have the asymptotic behavior: 
\eq{\label{asympfermDirac}
\Psi_{0,{q}}= \frac{1}{r}\left(\begin{array}{c}
c_1 \,e^{i E(t+r)}\\c_2\,e^{i E(t-r)}
\end{array}\right) 
\quad,\quad
\Psi_{0,{q}}=\begin{cases}\frac{1}{r}
\left(\begin{array}{c}
0\\e^{i E(t+r)}
\end{array}\right)&q=1\\\frac{1}{r}
\left(\begin{array}{c}e^{i E(t-r)}\\0
\end{array}\right)&q=-1
\end{cases}
} 
In particular, this demonstrates that there is no energy gap between the $j=0$ (i.e. s-wave) modes and the higher angular momentum modes. Notice that the $j=0$ modes are polarized so that the $q=1$ mode is in-going with positive helicity and the $q=-1$ mode is out-going with negative helicity. This implies that the boundary condition for the $j=0$ sector on the monopole will necessarily relate the $q=1$ modes to the $q=-1$ modes. 

Additionally, note that we can compute the net electric current flux from these asymptotic states and we find that for a single Weyl fermion of charge $q$:
\eq{\oint_{S^2_\infty} j^\mu \,\hat{r}_\mu d^2\Omega=
\oint_{S^2_\infty}q\,\bar\psi\bar\sigma^\mu\psi~ \hat{r}_\mu \,d^2\Omega=c^\dagger c-q (c_1^\dagger c_1-c_2^\dagger c_2)~,
}
where $c,c^\dagger$ are the mode operators for the $j=0$ state. 

We can now easily solve for the spectrum of the massive Dirac operator. If we consider a pair of charged fermions $\psi,\tilde\psi$ and 
turn on a mass term of the form 
\eq{
\CL_{\rm mass}=m_\psi \psi\tilde\psi+c.c.~,
}
where $m_\psi\in \IR$, then the simplified Dirac equation becomes 
\eq{
\left(\sigma^3\partial_r+\sigma^3\CD_E \right)\Psi_q =-im_\psi \bar{\tilde{\Psi}}_{-q} ~,\\
\left(\sigma^3\partial_r+\sigma^3\CD_{-E} \right)\bar{\tilde{\Psi}}_{-q} =im_\psi \Psi_q ~.
} 
Here, the Weyl fermions effectively pair up to form Dirac fermions $\Upsilon=(\Psi_\alpha,\bar{\tilde\Psi}^{\dot\alpha})$ where here we drop the notational dependence on $j,q$, leaving it implicit. 
Since the mass term is $r$-independent, it does not affect the $r\to 0$ limit -- it  only affects the $r\to \infty$ limit. Now, the asymptotic Dirac equation is modified to 
\eq{
\left(\sigma^3\otimes \mathds{1}\left(\partial_r+\frac{1}{r}\right)-iE\sigma^3\otimes \sigma^3+m_\psi\mathds{1}\otimes \sigma^2\right)\left(\begin{array}{c}\Psi\\\bar{\tilde{\Psi}}\end{array}\right)=0~.
}
Again, the asymptotic solutions are plane waves of the form 
\eq{
\Upsilon\underset{r\to \infty}{\longrightarrow} \frac{e^{ i (Et+\omega\sigma^3 r)}}{r}\Upsilon_{0,}~, 
}
where $\omega=\sqrt{E^2-m_\psi^2}$ and $\Upsilon=(\Psi,\bar{\tilde\Psi})$. Due to the mass term, the solutions for $\Upsilon_{0 }$ are linear combinations of $\Psi_{0 },\tilde\Psi_{0 }$:
\eq{\label{massiveIRsol}
&\Upsilon_{0 }=c_1\Upsilon_{0 }^{(1)}+c_2\Upsilon_{0 }^{(2)}+c_3\Upsilon_{0 }^{(3)}+c_4\Upsilon_{0 }^{(4)}
\\&=
c_1\left(\begin{array}{c}
-\sqrt{1+\gamma}\\0\\\frac{m_\psi}{E}\frac{\gamma}{\sqrt{1+\gamma}}\\0
\end{array}\right)
+c_2\left(\begin{array}{c}
\sqrt{1-\gamma}\\0\\\frac{m_\psi}{E}\frac{\gamma}{\sqrt{1-\gamma}}\\0
\end{array}\right)
+c_3\left(\begin{array}{c}0\\
\sqrt{1+\gamma}\\0\\\frac{m_\psi}{E}\frac{\gamma}{\sqrt{1+\gamma}}
\end{array}\right)
+c_4\left(\begin{array}{c}0\\
-\sqrt{1-\gamma}\\0\\\frac{m_\psi}{E}\frac{\gamma}{\sqrt{1-\gamma}}
\end{array}\right)
}
for $j>0$ and 
\eq{
\Upsilon_{0 }=b_1\Upsilon_{0 }^{(3)}+b_2\Upsilon_{0 }^{(4)}=b_1\left(\begin{array}{c}0\\
\sqrt{1+\gamma}\\0\\\frac{m_\psi}{E}\frac{\gamma}{\sqrt{1+\gamma}}
\end{array}\right)
+b_2\left(\begin{array}{c}0\\
-\sqrt{1-\gamma}\\0\\\frac{m_\psi}{E}\frac{\gamma}{\sqrt{1-\gamma}}
\end{array}\right)
}
for $j=0$ 
where $\gamma =\frac{1}{\sqrt{1-m_\psi^2/E^2}}$. 
Here the asymptotic behavior is given by 
\eq{
\Upsilon_{0 }&=c_1e^{i E(t+\omega r)}\Upsilon_{0 }^{(1)}+c_2e^{i E(t-\omega r)}\Upsilon_{0 }^{(2)}+c_3e^{i E(t-\omega r)}\Upsilon_{0 }^{(3)}+c_4e^{i E(t+\omega r)}\Upsilon_{0 }^{(4)}\\
&+b_1e^{i E(t-\omega r)}\Upsilon_{0 }^{(3)}+b_2e^{i E(t+\omega r)}\Upsilon_{0 }^{(4)}~,
} 
Using this mode expansion, we can compute the asymptotic charge flux:
\eq{\label{Upsiloncharges}\oint_{S^2_\infty} j^\mu \,\hat{r}_\mu d^2\Omega&=
\oint_{S^2_\infty} \left(\bar\psi_1\bar\sigma^\mu \psi_1+\psi_{-1}\sigma^\mu \bar{\psi}_{-1}\right)~\hat{r}_\mu\, d^2\Omega\\
&=-c_1^\dagger c_1-c_2^\dagger c_2+c_3^\dagger c_3+c_4^\dagger c_4+b_1^\dagger b_1+b_2^\dagger b_2~.
}

For the case where $E=m_\psi$, the solutions have a similar asymptotic structure as the case where $E,m_\psi=0$: there are always a pair of solutions that have polynomial  growth $r^{-1\pm \lambda}$ for $\lambda=\sqrt{j(j+1)}$ at $r\to \infty$. There are additionally a pair of solutions that grows like 
\eq{
\Upsilon \sim \begin{cases}\frac{r^\lambda}{\log(r)} &j>0\\
1&j=0
\end{cases}
} 
As we will see in the following section, the modes with this asymptotic behavior will not arise in our analysis due to the boundary conditions on the monopole and therefore we will not comment further on them. 

\subsection{Fermions and the Smooth Non-Abelian Monopole}

We will now discuss the boundary conditions on the IR modes. Since the dynamical dyon degree of freedom lives on the boundary, the fermion ``boundary conditions'' will include a boundary interaction with $\varphi(t)$. We can derive the interaction  by computing the mode expansion of the fermions in the fixed non-abelian monopole background \eqref{Connection} with $\varphi(t)\to 0$. Using this solution, we can expand the asymptotic form of the non-abelian fermion modes in terms of IR modes 
and then infer the coupling to $\varphi(t)$ by looking at the gauge orbit at $r\to 0$ since $\varphi(t)$ transforms as a Goldstone under $U(1)$ gauge transformations. Equivalently, we can solve for the mode expansion in the fixed monopole background with $\varphi(t)=0$ \eqref{Connection},  expand the solutions in terms of the IR modes, and then plug the result into the action with the full $\varphi(t)$-dependent connection \eqref{Adyon}.

Again, 
we can expand the fermion fields into representations of the angular momentum operator \eqref{Jnonabelian}. We will label these representations  by the same quantum numbers $j,m$.  
Here the angular momentum representations will have integer total spin $j\in  \IZ_{\geq 0}$ because the fundamental Weyl fermions transform in the spin-$\half$ representation of the $SU(2)$ gauge group. 

We are  interested in finding the spectrum of the massive Dirac operator for general total angular momentum $j$. Again, let us first find the spectrum of the massless Dirac operator for a single fermion   $\psi^{i}$ where $i=\pm$ is the gauge index.  

As before, we will simplify the Dirac equation with background gauge field \eqref{Connection} by performing a frame rotation by 
\eq{\label{framerotation}
\psi^i=U(\theta,\phi)\hat\psi^i\quad, \quad U(\theta,\phi) =e^{\frac{i \phi}{2}\sigma^0\bar\sigma^3}e^{\frac{i\theta}{2}\sigma^0\bar\sigma^2}~.
}
The resulting Dirac equation is given by   
\eq{
\Bigg[&-i E +\sigma^3\left(\partial_r+\frac{1}{r}\right)+\frac{\hat\CK}{r} 
-\left(\sigma^-e^{-i \phi}M_+-\sigma^+ e^{i \phi}M_-\right)\Bigg]\hat\psi^i=0~,
}
where $\sigma^\pm=\half(\sigma^1\pm i \sigma^2)$,  $M_\pm=T_\pm \,m_Wr\csch(m_Wr)$, and 
\eq{\label{ckdef}
\hat\CK&=
\sigma^1\left(\partial_\theta+\half \cot(\theta)\right)+\frac{\sigma^2}{\sin\theta}\Big((\partial_\phi+i T_3)-i \cos(\theta)T_3\Big)~, 
}
We can then expand $\hat\psi^\pm$ into angular momentum eigenstates as 
\eq{
\hat\psi^{+}= \left(\begin{array}{c}
f_{j,+} (r)\,e^{i \phi/2} \CD^{(j)}_{m,1}(\theta,\phi)\\
g_{j,+} (r)\,e^{-i \phi/2} \CD^{(j)}_{m,0}(\theta,\phi)
\end{array}\right)\quad, \quad 
\hat\psi^{- }= \left(\begin{array}{c}
f_{j,-} (r)\,e^{i \phi/2} \CD^{(j)}_{m,0}(\theta,\phi)\\
g_{j,-} (r)\,e^{-i \phi/2} \CD^{(j)}_{m,-1}(\theta,\phi)
\end{array}\right)~,
}
where   $\CD^{(j)}_{m,q}(\theta,\phi):=\CD^{(j)}_{m,q}(\phi,\theta,\phi)$ are Wigner $D$-functions. Again, when $j=0$, $\CD^{(0)}_{m,\pm1}$ is singular so $f_{j,+} ,g_{j,-} :=0$ and the Dirac equation further simplifies. 

Now, the Dirac equation reduces to a matrix ODE 
\eq{\label{finalDirac}
\left(\sigma^3\left(\partial_r+\frac{1}{r}\right)+\frac{1}{r}\left(\begin{array}{cc}
0&-\lambda\\
\lambda &0
\end{array}\right)-iE+\frac{1}{r}\left(\sigma^-M_+-\sigma^+M_-\right)\right)\Psi =0
}
 for the spin-$j$ eigen-spinors 
\eq{\label{Diracrspinor}
\Psi =\left(\begin{array}{c}
f _{j,+}(r)\\
g _{j,+}(r)\\
f _{j,-}(r)\\
g _{j,-}(r)
\end{array}\right)
}
where $\lambda=\sqrt{j(j+1)}$. Because the Dirac operator is everywhere smooth, it means that the solutions of the Dirac operator are formally given by solving the matrix ODE:
\eq{
\Psi=P\, e^{-\int \CD_E(r')\,dr'}\Psi_{0,{\rm UV}}\quad, \quad \CD_E(r)=\left(\begin{array}{cccc}
-i E&\frac{\lambda}{r}&&\\
\frac{\lambda}{r}&iE&\frac{1}{r}&\\
&\frac{1}{r}&-iE&\frac{\lambda}{r}\\
&&\frac{\lambda}{r}&iE
\end{array}\right)+\frac{1}{r}\mathds{1}~.
}
These formal solutions may however be non-normalizable.\footnote{Here the correct notion of the normalizable is plane-wave normalizable.} The normalizability of the solutions of this differential equation are now determined by the asymptotic behavior of the solutions at both $r\to\infty$ and $r\to 0$.

For $E=0$, we can explicitly solve for the solutions. 
As shown in \cite{Brennan:2021ucy}, {there are no normalizable solutions to the Dirac equation with $E=0$ and $j>0$}. However, for $j=0$, the solution is given explicitly by 
\eq{
\Psi=\frac{\tanh(m_Wr)}{r}\left(\begin{array}{c}
0\\1\\-1\\0
\end{array}\right)~. 
}
In the long distance limit, this solution is the combination of $E=0,$ $j=0$ modes of $\psi_q$ of opposite charges in the IR \eqref{DiracE0}:
\eq{
\Psi\underset{r\to \infty}{\longrightarrow} \left(\begin{array}{c}
\Psi_{+1}\\
-\Psi_{-1}
\end{array}\right)
}
 This will imply that the boundary condition at the monopole, is given by relating $\psi^{+}$ to $\psi^{-}$ via the dyon degree of freedom.

For $E>0$, the analysis is more involved. 
 At $r\to \infty$, the Dirac equation can be written as 
\eq{
\left(\sigma^3\left(\partial_r+\frac{1}{r}\right)+\frac{1}{r}\left(\begin{array}{cc}
0&-\lambda\\
\lambda &0
\end{array}\right)-iE\right)\Psi=0~. 
} 
The asymptotic behavior of the solutions match that of the IR: 
\eq{\label{asymptoticmassless}
\Psi\underset{r\to \infty}{\longrightarrow} \frac{e^{ i E(t-\sigma^3 r)}}{r}\Psi_{0,{\rm UV}}~. 
}
where $\Psi_{0,{\rm UV}}$ is a constant spinor. All solutions are asymptotically plane-wave normalizable for $E\neq 0$. 
Near $r\to 0$, the Dirac operator can be written as
\eq{\label{rtozeroDirac}
\left(\sigma^3\left(\partial_r+\frac{1}{r}\right)+\frac{1}{r}\left(\begin{array}{cc}
0&-\lambda\\
\lambda &0
\end{array}\right)+\frac{1}{r}\left(\sigma^-T_+-\sigma^+T_-\right)\right)\Psi=0~.
}
This operator has normalizable solutions which are given by 
\eq{
\Psi\sim c_1 \frac{(m_Wr)^{j+1}}{r}\left(\begin{array}{c}
\sqrt{j}\\-\sqrt{j+1}\\\sqrt{j+1}\\-\sqrt{j}
\end{array}\right)+c_2 \frac{(m_Wr)^{j}}{r}\left(\begin{array}{c}
\sqrt{j+1}\\-\sqrt{j}\\-\sqrt{j}\\\sqrt{j+1}
\end{array}\right)~,
}
for $j>0$ and 
\eq{
\Psi\sim \frac{1}{r}\left(\begin{array}{c}
0\\1\\-1\\0
\end{array}\right)~,
}
for $j=0$.

This means that the normalizable, positive energy solutions of the Dirac equation are those for which $E>0$ and $\Psi_{0,{\rm UV}}$ is given by 
\eq{\label{constferm}
\Psi_{0,{\rm UV}}=c_1\Psi_{0,{\rm UV}}^{(1)}+c_2\Psi_{0,{\rm UV}}^{(2)}=c_1 \left(\begin{array}{c}
\sqrt{j}\\-\sqrt{j+1}\\\sqrt{j+1}\\-\sqrt{j}
\end{array}\right)+c_2 \left(\begin{array}{c}
\sqrt{j+1}\\-\sqrt{j}\\-\sqrt{j}\\\sqrt{j+1}
\end{array}\right)~,
}
for $j>0$ and 
\eq{
\Psi_{0,{\rm UV}}=\Psi_{0,{\rm UV}}^{(0)}=
\left(\begin{array}{c}
0\\1\\-1\\0
\end{array}\right)}
for $j=0$. 
In particular, this demonstrates that the interaction with the monopole does not introduce an energy gap between the $j=0$ (i.e. s-wave) modes and the higher angular momentum modes. In terms of the IR modes \eqref{constfermIR}, these can be matched by 
\eq{
\Psi_{0,{\rm UV}} = c_1 \left(\begin{array}{c}
\sqrt{j}\Psi^{(1)}_{0,+1}-\sqrt{j+1}\Psi^{(2)}_{0,+1}\\\sqrt{j+1}\Psi^{(1)}_{0,-1}-\sqrt{j}\Psi^{(2)}_{0,-1}
\end{array}\right)+c_2\left(\begin{array}{c}
\sqrt{j+1}\Psi^{(1)}_{0,+1}-\sqrt{j}\Psi^{(2)}_{0,+1}\\-\sqrt{j}\Psi^{(1)}_{0,-1}+\sqrt{j+1}\Psi^{(2)}_{0,-1}
\end{array}\right)~, }
for $j>0$ and 
\eq{
\Psi_{0,{\rm UV}}=
\left(\begin{array}{c}
\Psi_{0,+1}\\-\Psi_{0,-1}
\end{array}\right)}
for $j=0$. It is important to note that these solutions all source an electric charge flux at infinity:
\eq{\oint_{S^2_\infty} j^\mu ~\hat{r}_\mu d^2\Omega=
\int_{S^2_\infty} \bar\psi_i\bar\sigma^\mu \psi^i\, \hat{r}_\mu\, d^2\Omega=2(c_1^\dagger c_1-c_2^\dagger c_2+c^\dagger c)~,
}
for fixed $E$ where $c_i$ are as in \eqref{constferm} and $c^\dagger,c$ are the creation and annihilation operators for the $j=0$ modes.

We can now easily solve for the spectrum of the massive Dirac operator. If we take two fermions, which we label as $\psi^i,\tilde\psi^i$, then 
turn on a mass term of the form 
\eq{
\CL_{\rm mass}=m_\psi \epsilon_{ij}\psi^i\tilde\psi^j+c.c.~,
}
where $m_\psi\in \IR$. With this deformation, the simplified Dirac equation reduces to 
\eq{
\left(\sigma^3\partial_r+\sigma^3\CD_E \right)\Psi=-im_\psi \bar{\tilde{\Psi}}~,\\
\left(\sigma^3\partial_r+\sigma^3\CD_{-E} \right)\bar{\tilde{\Psi}}=im_\psi \Psi~,
}
where $\Psi,\tilde\Psi$ are the analogs of the $r$-dependent spinors in  \eqref{Diracrspinor} for $\psi,\tilde\psi$ respectively. 
Here, the fermions again effectively pair up to form Dirac fermions $\Upsilon=(\Psi_\alpha,\bar{\tilde\Psi}^{\dot\alpha})$. 
Since the mass term is $r$-independent, it does not affect the $r\to 0$ limit. However, it does affect the $r\to \infty$ limit. Now, the asymptotic behavior is modified to 
\eq{\label{asymptoticmassive}
\Upsilon\sim \frac{1}{r}e^{\pm i \omega r}\Upsilon_{0,{\rm UV}}~,
}
for $E\neq m_\psi$ where $\omega=\sqrt{E^2-m_\psi^2}$. Here, the constant Dirac fermion $\Upsilon_{0,{\rm UV}}$ takes the form
\eq{\label{asymptoticmassivetotal1}
\Upsilon_{0,{\rm UV}}&=a_1\left(\begin{array}{c}
\Psi_{0,{\rm UV}}^{(1)}\\-\frac{m_\psi}{E}\frac{\gamma}{1+\gamma}\bar\Psi_{0,{\rm UV}}^{(1)}
\end{array}\right)+a_2\left(\begin{array}{c}
\Psi_{0,{\rm UV}}^{(1)}\\\frac{m_\psi}{E}\frac{\gamma}{1-\gamma}\bar\Psi_{0,{\rm UV}}^{(1)}
\end{array}\right)\\
&+a_3\left(\begin{array}{c}
\Psi_{0,{\rm UV}}^{(2)}\\-\frac{m_\psi}{E}\frac{\gamma}{1+\gamma}\bar\Psi_{0,{\rm UV}}^{(2)}
\end{array}\right)+a_4\left(\begin{array}{c}
\Psi_{0,{\rm UV}}^{(2)}\\\frac{m_\psi}{E}\frac{\gamma}{1-\gamma}\bar\Psi_{0,{\rm UV}}^{(2)}
\end{array}\right)
}
for $j>0$ and 
\eq{
\Upsilon_{0,{\rm UV}}=b_1\left(\begin{array}{c}\Psi_{0,{\rm UV}}^{(0)}\\ -\frac{m_\psi}{E}\frac{\gamma}{1+\gamma}\bar\Psi_{0,{\rm UV}}^{(0)}\end{array}\right)
+b_2\left(\begin{array}{c}\Psi_{0,{\rm UV}}^{(0)}\\ \frac{m_\psi}{E}\frac{\gamma}{1-\gamma}\bar\Psi_{0,{\rm UV}}^{(0)}\end{array}\right)~,
}
for $j=0$, where again $\gamma=\frac{1}{\sqrt{1-m_\psi^2/E^2}}$ 
. 

The asymptotic behavior of these solutions is given by 
\eq{\label{asymptoticmassivetotal2}
\Upsilon&\sim \frac{e^{i (Et+\omega \sigma^3 r)}}{r}\left\{a_1\left(\begin{array}{c}
\Psi_{0,{\rm UV}}^{(1)}\\-\frac{m_\psi}{E}\frac{1}{1+\gamma}\bar\Psi_{0,{\rm UV}}^{(1)}
\end{array}\right)
+a_3\left(\begin{array}{c}
\Psi_{0,{\rm UV}}^{(2)}\\-\frac{m_\psi}{E}\frac{1}{1+\gamma}\bar\Psi_{0,{\rm UV}}^{(2)}
\end{array}\right)\right\}\\
&+\frac{e^{i (Et-\omega \sigma^3 r)}}{r}\left\{a_2\left(\begin{array}{c}
\Psi_{0,{\rm UV}}^{(1)}\\\frac{m_\psi}{E}\frac{\gamma}{1-\gamma}\bar\Psi_{0,{\rm UV}}^{(1)}
\end{array}\right)
+a_4\left(\begin{array}{c}
\Psi_{0,{\rm UV}}^{(2)}\\\frac{m_\psi}{E}\frac{\gamma}{1-\gamma}\bar\Psi_{0,{\rm UV}}^{(2)}
\end{array}\right)\right\}~,
}
for $j>0$ and 
\eq{
\Upsilon&\sim \frac{e^{i (Et+\omega \sigma^3 r)}}{r}b_1\left(\begin{array}{c}\Psi_{0,{\rm UV}}^{(0)}\\ -\frac{m_\psi}{E}\frac{\gamma}{1+\gamma}\bar\Psi_{0,{\rm UV}}^{(0)}\end{array}\right)
+\frac{e^{i (Et-\omega \sigma^3 r)}}{r}b_2\left(\begin{array}{c}\Psi_{0,{\rm UV}}^{(0)}\\ \frac{m_\psi}{E}\frac{\gamma}{1-\gamma}\bar\Psi_{0,{\rm UV}}^{(0)}\end{array}\right)~,
}
for $j=0$. 
Again, these can be expanded in terms of the IR modes \eqref{massiveIRsol}. For brevity, we will not give the full expressions. However, it is clear by comparing the form of the $\Upsilon_{0,{\rm IR}}^{(a)}$ and their associated charges in \eqref{massiveIRsol} -- \eqref{Upsiloncharges} and form of the $\Upsilon_{0,{\rm UV}}$ in \eqref{asymptoticmassivetotal1}, that these solutions relate an IR mode with positive asymptotic charge flux to and IR mode with negative asymptotic charge flux. However, these combinations are such that the total asymptotic flux is given by 
\eq{\label{TotalModeFlux}\oint_{S^2_\infty} j^\mu ~\hat{r}_\mu d^2\Omega=
\oint_{S^2_\infty}\bar\Upsilon \gamma^\mu \Upsilon\, \hat{r}_\mu\, d^2\Omega=2(a_1^\dagger a_1+a_2^\dagger a_2-a_3^\dagger a_3-a_4^\dagger a_4)+2(b_1^\dagger b_1+b_2^\dagger b_2)~.
}

For the case where $E=m_\psi$, the solutions have a similar asymptotic structure as the case where $E,m_\psi=0$: there is a $j=0$ mode that has $1/r$ behavior at $r\to \infty$ at energy $E=m_\psi$ but no normalizable $j>0$ modes with energy $E=m_\psi$.

Additionally, one can show by explicitly checking, that solutions with $|E|<|m_\psi|$ are not normalizable because the $\Upsilon_{0,{\rm UV}}$ that lead to normalizable solutions in the $r\to0$ limit turn on the exponential growth at $r\to \infty$. Physically, this reflects the fact that turning on masses induces an energy gap. 
Therefore, there is again a continuous spectrum of states for massive fermions which has the same behavior for $E\geq m_\psi$ as the spectrum of states for massless fermions at energies $E\geq0$.

\section{Generalized Symmetries and the Callan Rubakov Effect}
\label{sec:symmetries}

Now let us discuss the symmetries of $4d$ $SU(2)$ gauge theory with $N_f$ {Weyl fermions $\psi^{iA}$ that transform in the fundamental representation of the gauge group where $i=\pm$ are $SU(2)$ indices and $A=1,...,N_f$}. We will take $N_f=2n_f$ so that the theory has no gauge anomaly {and we will suppress spinor indices, implicitly using the spinor-index contraction convention as in \cite{Wess:1992cp}}. Here we will use the language of generalized global symmetries;  for a review see \cite{Brennan:2023mmt,Schafer-Nameki:2023jdn,Bhardwaj:2023kri,Shao:2023gho}.

Let us first consider the symmetries of the massless theory. This theory is described by the action 
\eq{
S=\int \frac{1}{2g^2}\Tr[F\wedge \ast F]+i \bar\psi_{iA}\slashed{D}\psi^{iA}
+ \Tr|D\Phi|^2-V(\Phi)~,
}
where again we choose a potential $V(\Phi)$ so that $\Phi$ condenses above the confinement scale. This theory has a $SU(N_f)^{(0)}$ 0-form global symmetry that rotates the fermions which we will often refer to as the 0-form \textit{flavor symmetry}.   Note that there is no independent $U(1)$ chiral symmetry because it is broken to the center of $SU(N_f)^{(0)}$ by an ABJ anomaly. 

In the IR the theory flows to $U(1)$ gauge theory with $N_f$ {Weyl fermions of positive unit electric charge $\psi^A=\psi^{+A}$ and $N_f$ Weyl fermioins of negative unit electric charge $\tilde\psi^A=\psi^{-A}$}. 
Here we see in the IR that the theory has a $SU(N_f)^{(0)}\times SU(N_f)^{(0)}$ global flavor symmetry where each factor acts only on  $\psi^A$ or $\tilde\psi^A$ in the fundamental representation. 
Additionally, this theory has a $U(1)^{(1)}$ 1-form magnetic symmetry which protects the monopole lines from decay. \footnote{Note that in this theory there is a mixed $(SU(N_f)^{(0)}\times SU(\Nf)^{(0)})^2$-$U(1)$ mixed anomaly which gives rise to a 2-group global symmetry involving the magnetic 1-form global symmetry \cite{vanBeest:2023dbu}. This 2-group global symmetry is broken by fermion mass terms and therefore will not play a role in our analysis. 
}

The theory also develops a $\IZ_{N_f}$ generalized symmetry structure which we denote as $\IZ_\Nf^{(0,1)}$. 
This follows from the fact that the fermions transform faithfully under \cite{Brennan:2022tyl,Delmastro:2022pfo,Hsin:2020nts,Lee:2021crt} 
\eq{\label{1formUV}
G_{\rm ferm} =\frac{U(1)_{\rm gauge}\times SU(N_f)^{(0)}\times SU(N_f)^{(0)}}{\IZ_{N_f}}~. 
}
We would like to elaborate on the nature of this  generalized global symmetry structure.\footnote{We would like to especially thank Konstantinos Roumpedakis and Pierluigi Niro for related discussions.}

A 1-form center symmetry typically arises in $G$ gauge theories when there is matter that is uncharged under a subgroup $\Gamma$ of the center of the gauge group: $\Gamma\subset Z(G)$. When $\Gamma$ is non-trivial, we can restrict the path integral to only integrate over $G/\Gamma$ gauge fields with a fixed, non-trivial discrete flux $w_2(\Gamma)\in H^2(BG/\Gamma,\Gamma)$ that obstructs lifting the $G/\Gamma$ gauge field  to a $G$ gauge field. The action of the 1-form $\Gamma^{(1)}$-symmetry is implemented by a $\Gamma$-valued large gauge transformation of the $G/\Gamma$ gauge field which corresponds to a shift in the choice of integer lift of $w_2(\Gamma)$. This large gauge transformation acts non-trivially on  Wilson lines $W_R$ of representations $R$ that transform non-trivially under $\Gamma\subset Z(G)$ because it is not a $G$ gauge transformation, but a discrete shift in the space of gauge $G$ gauge fields. This action multiplies $W_R(\gamma)\mapsto W_R(\gamma)\,e^{i n \oint_\gamma \Lambda_1}$ where $\Lambda_1$ is the $\Gamma$-valued transformation parameter. This symmetry can be enacted by a co-dimension 2 topological symmetry defect operator (SDO) which induces a large gauge transformation on a Wilson line when the defect crosses the line. Physically, the SDO operator can be viewed as measuring the electric flux through the surface, or conversely as a topological flux tube that induces a $\Gamma$-valued Aharanov-Bohm phase along any linking circle. 

In our setting, the fields are invariant under a diagonal combination of $\IZ_{N_f}\subset Z(U(1))$ and $\IZ_{N_f}\subset Z\Big(SU(N_f)\times SU(N_f)\Big)$. This means that the theory can be described by $U(1)$ gauge theory with a $G_f/\IZ_{N_f}$ global symmetry where $G_f=SU(N_f)\times SU(N_f)$. This theory admits all $U(1)$ Wilson lines and we are allowed to dress any line operator by $G_f/\IZ_{N_f}$ background Wilson lines -- i.e. we are allowed to add local counter terms which are dependent on the $G_f/\IZ_{N_f}$ background gauge connection.\footnote{Since the global symmetry is given by $G_f/\IZ_{N_f}$, it is a consistent choice to only allow local counter terms that are dependent on the $G_f/\IZ_{N_f}$ connection (and not a $G_f$ connection). 
}

As in the case of standard 1-form center symmetries, we can restrict the path integral to $U(1)/\IZ_{N_f}$ gauge fields with fixed discrete flux $w_2(\IZ_{N_f})$ if we additionally couple the theory to a $G_f/\IZ_\Nf$ 0-form background gauge field with the same discrete flux. The path integral is then invariant under simultaneous large $\IZ_\Nf$ gauge transformations of the  $U(1)/\IZ_\Nf$ and $G_f/\IZ_\Nf$ gauge fields (up to possible anomalies):
\eq{
Z[w_2(\IZ_\Nf)+N_f\Lambda_2]=Z[w_2(\IZ_\Nf)]~. 
}
The invariance of the path integral under this shift indicates a $\IZ_\Nf$ symmetry which appears similar to a 1-form global symmetry due to the fact that the matter fields are invariant under the diagonal $\IZ_{N_f}\subset U(1)\times G_f$. Indeed, the simultaneous $\IZ_\Nf$ large gauge transformations of the $U(1)/\IZ_{N_f}$ dynamical and $G_f/\IZ_\Nf$ background gauge fields acts on the Wilson lines as $W_q(\gamma)\mapsto W_q(\gamma)\, e^{i q \int_\gamma \Lambda_1}$ where $\Lambda_1$ is the large $\IZ_{N_f}$ gauge transformation parameter.

However, there are several subtleties that arise in trying to understand in what way the  invariance of the path integral under the large $\IZ_\Nf$ gauge transformations is a symmetry.  The first that arises is how one would construct the topological operator that enacts the symmetry since the gauge field for $G_f/\IZ_\Nf$ is not an operator in our theory. A symmetry defect operator for this ``symmetry'' must induce large $\IZ_\Nf$ transformations in $G_f/\IZ_\Nf\times U(1)/\IZ_\Nf$. The shift of the $U(1)/\IZ_\Nf$ field is enacted by the (broken) 1-form center symmetry generator. Similarly, the shift operator for $G_f/\IZ_\Nf$ is generated by a flux operator which can be thought of as {an Aharanov-Bohm flux tube/Gukov-Witten Operator \cite{Gukov:2006jk,Gukov:2008sn} in $G_f/\IZ_\Nf$ around which the $G_f/\IZ_\Nf$ has a holonomy that winds around a non-trivial element of $\pi_1(G_f/\IZ_\Nf)$. }   Since $G_f$ is simply connected, this winding is indeed topologically non-trivial (i.e. $\pi_1(G_f/\IZ_\Nf)\neq 0$) and the existence of the flux operator is guaranteed by the fact that we are allowed to turn on any $G_f/\IZ_\Nf$-bundle up to the matching of discrete flux. Similar constructions were considered in \cite{Hsin:2022heo}. 

If we take the product of these two non-topological operators, we construct a topological operator $\CU_g(\Sigma)$. This operator is topological since it generates a non-trivial flux associated to the Poincar\'e dual of $\Sigma$, \footnote{Here $[\Sigma]^\vee$ is the 2-form which is  Poincar\'e dual to $\Sigma\in H_2(M;\IZ)$. This is defined by the relation $\oint_\Sigma [\Sigma]^\vee=1$. } 
\eq{
\CU_g(\Sigma)~\longmapsto ~w_2(\IZ_\Nf)=n\cdot  [\Sigma]^\vee \quad, \quad g=e^{\frac{2\pi i n}{\Nf}}~.
}
Since a  deformation of $\Sigma$ acts trivially on its topological class, a deformation of the surface $\Sigma$ will at most induce a gauge transformation of $w_2(\IZ_\Nf)$:
\eq{
\delta_\Sigma \CU_g(\Sigma)~\longmapsto~\delta w_2(\IZ_\Nf)=[\Sigma+\delta \Sigma]^\vee-[\Sigma]^\vee=
[0]\in H^2(M;\IZ_{N_f})~. 
}
The invariance of the action under gauge transformations of $w_2(\IZ_\Nf)$  along the diagonal embedding of $U(1)/\IZ_\Nf\times G_f/\IZ_\Nf$, implies that any correlation function does not depend on $\Sigma$ except through its topological class and therefore that the associated operator is topological.

\begin{figure}
\begin{center}
\includegraphics[scale=0.45,clip,trim=3cm 15cm 3cm 1cm]{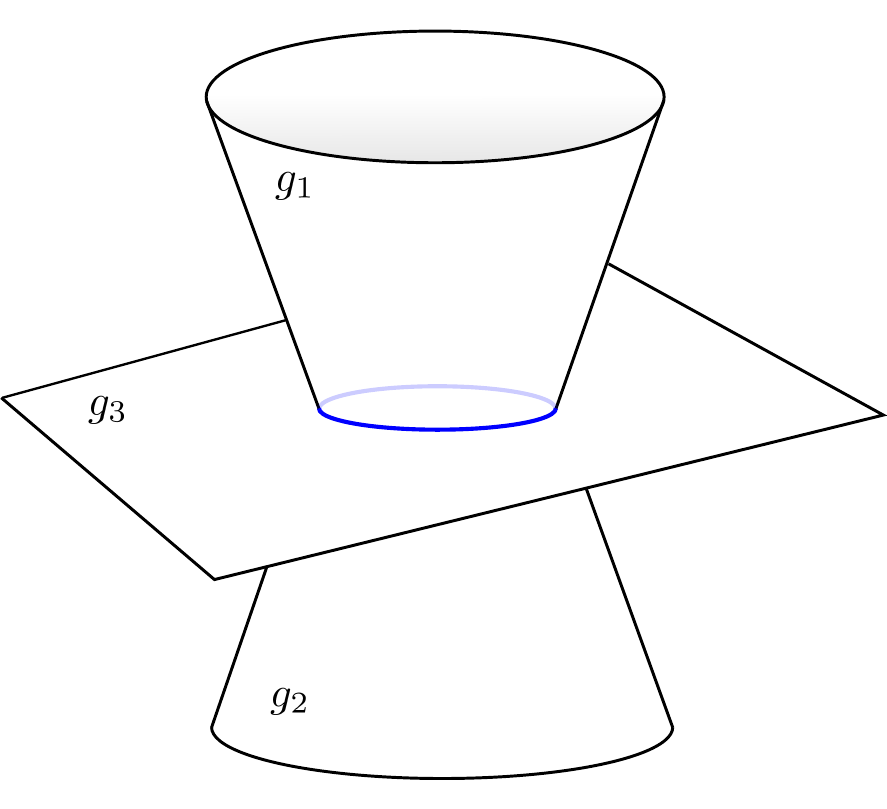}
\includegraphics[scale=0.6,clip,trim=5cm 4cm 8cm 15.5cm]{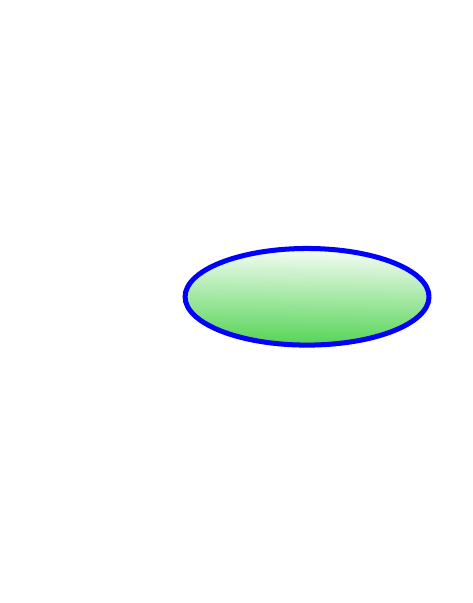}
\end{center}
\caption{In this figure (left) we illustrate how $\IZ_\Nf $ symmetry defect operator (blue) must come with a choice of ``invisible'' 0-form $G_f/\IZ_\Nf$ symmetry defect operators (black) in $3d$.  Here, $g_1,g_2,g_3\in G_f/\IZ_\Nf$ and $g_1g_2g_3^{-1}\in \IZ_\Nf$ such that the $G_f/\IZ_\Nf$ winding is implemented by the 0-form SDOs. On the right, we give an alternative construction where the symmetry defect operator (blue) wraps a $S^1$ is dressed by 0-form SDOs (green) on $D^2$.  \label{fig:1formSDO}}
\end{figure}

\bigskip
 This defect acts on Wilson lines by linking: pulling the defect through the Wilson line operator  induces a large $\IZ_\Nf$ gauge transformation on the Wilson line similar to an ordinary 1-form global symmetry.  
However, since the fermions have charge $\pm1$ in our theory, the Wilson lines can be cut by fermion operators. Despite this, the $\IZ_\Nf$ symmetry is preserved due to the fact that  the end points of a cut Wilson line transform under a projective $G_f/\IZ_{N_f}$ representation. {Because of this, we must also specify the lift of the $G_f/\IZ_\Nf$-bundle to a $G_f$-bundle in defining the topological operator. In general, it is not possible to lift a generic $G_f/\IZ_{N_f}$-bundle to a $G_f$-bundle without introducing branch } singularities which we can interpret as Aharanov-Bohm strings. However, if we consider locally the deformation of the bundle due to the insertion of $\CU_g(\Sigma)$, then there is locally a  lift when $\Sigma$ is contractible.\footnote{The reason is that if $\Sigma$ is contractible, the obstruction to lift the $G_f/\IZ_\Nf$-bundle to a $G_f$-bundle is trivializable. } This lift requires making a choice of $\IZ_\Nf$ 0-form symmetry defect operators that end on $\Sigma$ and implement the $\IZ_\Nf$ winding of the associated $G_f$ 0-form gauge field on circles that link $\Sigma$. Such as in Figure \ref{fig:1formSDO}.   In other words, $\CU_g$ should come with a choice of dressing by $\IZ_\Nf\subset G_f$ symmetry defect operators that end on $\CU_g$ which, due to the identification of $\IZ_\Nf\subset U(1)\sim \IZ_\Nf\subset G_f$, do not act on gauge invariant local operators.

\begin{figure}
\begin{center}
\includegraphics[scale=0.7,clip,trim=2cm 19cm 2cm 2cm]{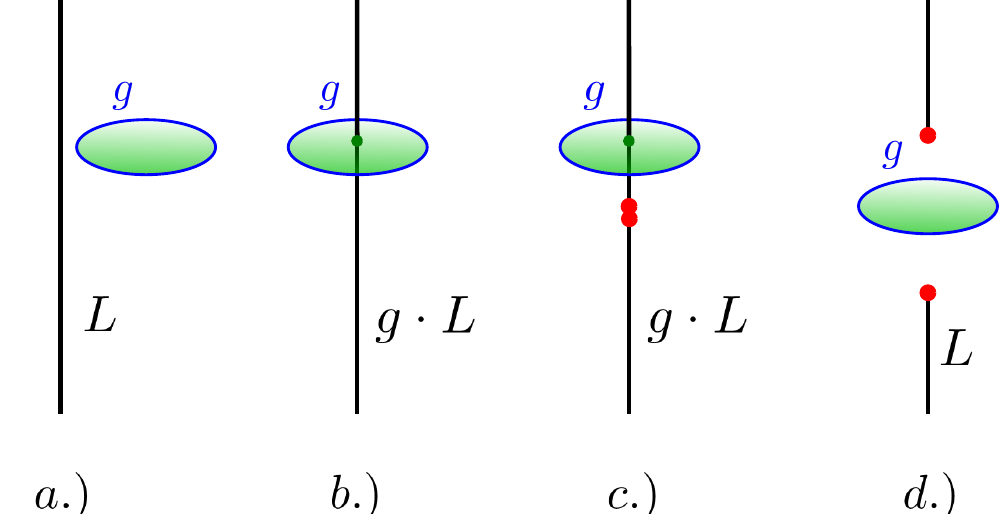}
\end{center}
\caption{In this figure we illustrate how a Wilson line  in a theory that couples to $\frac{G_{\rm gauge}\times G_{f}}{\IZ_\Nf}$-bundles   can be charged under the $\IZ_\Nf$ symmetry even though it  can be cut by a local operator.  The un-cut line operator $(a.)$ is acted on $(b.)$ by the $g$-symmetry defect operator which is constructed by stretching 0-form (green) SDO across a 1-form symmetry defect (blue). The line operator can then be cut by inserting a pair of charged local operators $(c.)$. The string can be un-linked from $\CU_q$ by pulling the end-points through the 0-form defect which is the inverse of the linking action $(d.)$. Here the  symmetry acts as a phase in correlation functions due to the fact that $\CU_q$ implements a singular large gauge transformation.  \label{fig:endpoints}}
\end{figure}

Now, if we try to unlink any charged Wilson line from the symmetry defect operator $\CU_q$ by cutting with charged point operators, the end points will see the $\IZ_\Nf\subset G_f$ 0-form symmetry defect operators attached to $\CU_q$. The action of these $\IZ_\Nf$ 0-form SDOs will implement the $\IZ_\Nf$ action on the cut line configuration 
see Figure \ref{fig:endpoints} so that the $\IZ_\Nf$ charge is preserved. 
Additionally, the charge of the fermions does not forbid them to dynamically screen the Wilson lines. However, due to the representation of the fermions, this will lead to a line operator that has $G_f/\IZ_\Nf$ symmetry fractionalization \cite{Delmastro:2022pfo,Brennan:2022tyl,Barkeshli:2014cna,Chen:2014wse,Brennan:2023ynm} -- i.e. a $G_f/\IZ_\Nf$ world volume anomaly  -- and again will be a line operator that is charged under $\IZ_\Nf$ symmetry.

We can turn on a background gauge field for this symmetry (i.e. turn on discrete flux associated to the quotient) by wrapping the symmetry defect operator on a non-trivial 2-cycle $\Sigma$, which, upon choosing a consistent set of $G_f/\IZ_\Nf$ 0-form SDOs, is equivalent to a choice of $G_f/\IZ_\Nf$-bundle with discrete flux $w_2(\IZ_\Nf)=[\Sigma]^\vee$  in addition to turning on a correlated discrete flux in the gauge group.  
However, this symmetry is un-gaugable without also gauging all of $G_f$ because summing over $w_2(\IZ_\Nf)$ obstruction classes/discrete fluxes requires making a choice of $G_f/\IZ_\Nf$-bundle for each flux sector. For this reason, we will call such a symmetry an \textit{intrinsically non-gaugable symmetry}.\footnote{{Note that this intrinsically non-gaugable symmetry is qualitatively different than a ``normal''  global symmetry with an `t Hooft anomaly. In the case of a symmetry with `t Hooft anomaly, the symmetry cannot be gauged because the partition function is not invariant under the gauge transformations whereas the intrinsically non-gaugable $\IZ_N^{(0,1)}$ symmetry structure cannot be gauged without first gauging the associated 0-form global symmetry. }
}

We anticipate that this construction of a topological operator by dressing a non-topological co-dimension $q$ defect operator, that is associated to a broken symmetry, by a non-genuine, non-topological operator that lives at the junction of $p<q$-form symmetry defects is much more general. As such, we will refer to such symmetries as \textit{$(p,q)$-form global symmetries} and denote them as $G^{(p,q)}$. In our setting, we are discussing a $(0,1)$-form symmetry, hence the notation $\IZ_\Nf^{(0,1)}$. We plan to further explore such symmetries in future investigations. 

\bigskip

\bigskip
Now let us study the symmetries of the theory with massive fermions. Turning on a mass breaks the flavor symmetry to (a subgroup of) $SU(N_f)^{(0)}\to SO(N_f)^{(0)}$. We will consider the case where  the fermions have an $SU(n_f)^{(0)}\subset SO(N_f)^{(0)}$ preserving mass 
\eq{
\CL_{\rm mass}=m_\psi \epsilon_{ij}\psi^{ia}\tilde\psi^j_a+c.c. 
}
$m_\psi\in \IR$ and $i,j=\pm$ are $SU(2)$ indices and $a=1,...,n_f$ are $SU(n_f)^{(0)}$ indices where here we have artificially divided the $N_f$ Weyl fermions into two sets of $n_f$: $\psi^{ia},\tilde\psi^i_a$ that transform in the fundamental and anti-fundamental representations of $SU(n_f)^{(0)}$ respectively.\footnote{Here we take $m_\psi\in \IR_{>0}$  so that we have a $CP$-preserving mass.}

In the IR the theory flows to massive $U(1)$ gauge theory with $2N_f$ fermions of mass $m_\psi$. 
We can then rewrite this in terms of $N_f$ fermions {of positive and negative unit electric charge: }
\eq{\label{psisplitting}
\psi^A=\left(\psi^{+a},\tilde\psi^+_a\right)\quad, \quad \tilde\psi_A=\left(\tilde\psi_a^-, \psi^{-a}\right)\quad, \quad A=1,...,N_f=2n_f~,
}
so that we can write the mass term as 
\eq{
\CL_{\rm mass}=m_\psi \left(\psi^{+a}\tilde\psi^-_a+\tilde\psi^+_a\psi^{-a}\right)+c.c.=m_\psi \psi^A\tilde\psi_A+c.c.~. 
}
Here we see in the IR that the theory has a $SU(N_f)^{(0)}$ global symmetry that acts on these $\psi^A,\tilde\psi_A$. Here we have organized the fermions based on their electric charge (as in the massless case) so that the positively (negatively) charged fermions transform in the fundamental (anti-fundamental) representation of $SU(N_f)^{(0)}$. This means that the symmetry group that acts faithfully is 
\eq{
G_{\rm ferm}=\frac{U(1)_{\rm gauge}\times SU(N_f)^{(0)}}{\IZ_{N_f}}~.
}
Again, the theory has a $U(1)^{(1)}$ 1-form magnetic symmetry and the quotient implies an additional emergent $\IZ_{N_f}^{(0,1)}$ (0,1)-form global symmetry at energies below $E \sim m_W$. 

\subsection{Constraints on Monopole-Fermion Scattering from Symmetries}
Now we will discuss the constraints that the conserved global symmetries impose on monopole-fermion scattering. Here we consider scattering fermions at energies $E\ll m_W$ so that the entire scattering process can be described in the low energy effective QED theory.  
The standard description of the Callan Rubakov effect is that an incident fermion will interact with the dyon degree of freedom due to the polarization of the s-waves in the fermion mode expansion \eqref{asympfermDirac}. Then, by conservation of angular momentum and gauge symmetry, fermion scattering in the $j=0$ sector leads to charge accumulation on the monopole which is then radiated away  due to the heavy mass of $\varphi(t)$. 
For the moment, we will assume that charge accumulation is a general property of monopole-fermion scattering and discuss what the generalized symmetries imply.

Consider scattering a charged fermion off of a monopole. 
An incident fermion will generically reach the monopole core. Besides examining the explicit fermion mode expansion, this follows from classical physics because a point particle can run into the monopole core if it approaches along the radial magnetic field lines due to the lack of a Lorentz force. 
Inside the monopole core, the scalar vev vanishes and $SU(2)$ gauge symmetry is restored. As the fermion scatters, it interacts with the dyon degree of freedom via the $SU(2)$ gauge interaction with the $W$-boson. Momentarily forgetting about energetics, this excites a $W$-boson and transforms the monopole into a dyon of charge 2. 

 {Now we can use the fact that there is a $\IZ_\Nf^{(0,1)}$ symmetry structure in this theory. It is important to note that this IR symmetry is broken at energy scales $E\sim m_W$ by the interaction with the massive $W$-bosons. Since the monopoles have trapped $W$-bosons due to $SU(2)$ gauge symmetry restoration in the core, the $\IZ_\Nf^{(0,1)}$ symmetry is locally broken on the monopole by $W$-boson excitations. However, such a local breaking of a symmetry is the statement that the dyonic operators are charged under the symmetry.   }

Now, the fact that there is a (0,1)-form $\IZ_{N_f}^{(0,1)}$ global symmetry, which acts on electrically charged line operators, implies that electric charge of the dyon cannot be radiated away without breaking the 0-form flavor symmetry.
This forbids the creation of asymptotic states with fractional fermion number and contradicts the original analysis of the Callan Rubakov effect which did not make use of this additional symmetry.

If the dyon charge cannot dissipate, then the scattered fermion is propagating in the presence of an electrically charged line operator. This changes the relevant mode quantization and  allows for a fermion bound state if the scattered fermion does not have enough energy to overcome the electrostatic potential.

In the case where we scatter $n_f=\frac{N_f}{2}$ fermions there is $N_f$ charge accumulated and  the $\IZ_{N_f}^{(0,1)}$ symmetry does not protect the dyon degree of freedom from decaying. Here, the dyon will decay into bulk, s-wave fermion modes by distributing the electric charge among the $N_f$ out-going s-wave modes. Here, we can see this from the fact that integrating out the dyon degree of freedom gives a local `t Hooft operator for the minimal $SU(2)$ instanton \cite{Fan:2021ntg}. In our theory, the fact that there are $N_f$ flavors means that there is an $N_f$-fermion local operator through which the dyon degree of freedom can decay.

We would like to emphasize that since we are scattering at energies $E\ll m_W$, that the $\IZ_{N_f}^{(0,1)}$ (0,1)-form global symmetry is not violated at long distances. This simple analysis from symmetries suggests that monopole-fermion scattering behaves qualitatively different than previous proposals.

\bigskip
To complete the analysis of monopole-fermion scattering from symmetry constraints,  we must now discuss how our proposal is consistent with conservation of energy. Naively, the energy required to excite the monopole into a dyon is on the order of the cutoff scale $E\sim m_W$. This can be seen from a number of arguments, but the simplest is probably to just note that the dyon is a monopole with an excited gauge boson which has mass $m_W$. However, not all dyons are constructed in this way. 

Let us recall the Witten effect \cite{Witten:1979ey}. Here we can take $U(1)$ Maxwell theory with a $4d$ $\theta$-angle:
\eq{
S=\int \frac{1}{2g^2}F\wedge \ast F+\frac{i\theta_{4d}}{8\pi^2}\int F\wedge F~.
}
As discussed in \cite{Witten:1979ey}, a non-zero $\theta_{4d}$ induces an electric charge on the monopole. Classically, this occurs because the $\theta$-term can be written as 
\eq{
\frac{i\theta_{4d}}{8\pi^2}\int F\wedge F=\theta_{4d} \int  \vec{E}\cdot \vec{B}~.
}
Since the magnetic monopole acts as a source for $B$, the interaction above acts to source an electric charge $q_e=\frac{\theta_{4d}}{2\pi} q_m$. In particular, shifting $\theta$ by $4\pi$ shifts the electric charge of the minimal monopole by $+2$. 

In the quantum theory, a $\theta$-angle can be generated by integrating out heavy fermions that have a $CP$-violating mass term $m_\psi=M_\psi e^{i \theta_{4d}}$ for $M_\psi>0$. Here, the electric charge is generated by the fermion vacuum which can be computed by summing over the contribution to the (renormalized) charge density from the continuum of negative energy  states in the Dirac sea. The vacuum charge density was computed in \cite{Yamagishi:1982wp,Grossman:1983yf} and results in the effective charge density: 
\eq{
\rho(r)=-\frac{M_\psi\,\sin(\theta_{4d})}{\pi}\int_{M_\psi}^\infty \frac{k\,e^{- 2k r} dk}{\sqrt{k^2-M_\psi^2}(k+M_\psi \cos(\theta_{4d}))}~.
}
Here the integrand is dominated by $k$ near $M_\psi$ so that the charge density is localized around the monopole with a radius of order $R_c\sim 1/|m_\psi|$. This contributes to the mass of the dyon on the order of $|m_\psi|$ due to the energy in the electric field. 

These energetics are also reflected in the quantum mechanics governing the dyon degree of freedom. In the monopole background, we can spherically reduce:
\eq{
\frac{i\theta_{4d}}{2\pi}\int \frac{F\wedge F}{4\pi}=\frac{i \theta_{4d}}{2\pi}\int F_{tr}\,dtdr=\frac{i\theta_{4d}}{4\pi}\int (\dot\varphi-2a_0) \,dt~,
}
so that the $4d$ $\theta$-term induces a $1d$ $\theta$-term with $\theta_{1d}=\frac{\theta_{4d}}{2}$ in the dyon quantum mechanics where in the final step  we inferred the coupling to $\varphi$ by gauge invariance. As we have discussed above, shifting $\theta_{4d}$ from 0 to $4\pi$ shifts $\theta_{1d}$ by $2\pi$ so that the ground state of the quantum mechanics shifts from the momentum eigenstate $|0\rangle\mapsto |1\rangle$ which sources a bulk electric charge field of charge 2. We would like to emphasize that these effects are not simply correlated, but rather are manifestations of the same physical effect as they both can be understood as coming from a shift in the $4d$ fermionic vacuum which couples to the gauge field as  $\theta_{4d}$. 

Together the constraints from gauge symmetry and angular momentum and from $\IZ_{N_f}^{(0,1)}$ symmetry imply that monopole scattering, at least in the $j=0$ sector,  leads to charge accumulation on the monopole which cannot be radiated away unless the total charge is a multiple of $N_f$. This suggests, due to the fact that the mass of the dyon degree of freedom is given by $m_W$, that the charge accumulation should be possible at low energy. 
Indeed, as we will discuss in detail in the following sections, the phase mode of the fermion acts like an axion field so that propagating charged fermions dynamically rotate the effective $\theta$-angle so that the gauge charge is conserved at low energies by a dynamical version of the Witten effect. 
For example, we will show that in s-wave scattering, where a positively charged in-going fermion scatters into a negatively charged out-going fermion, there is an effective $4\pi$ $\theta$-angle rotation on the monopole that accounts for the conservation of gauge charge. 
One of the main goals of the rest of the paper to demonstrate this dynamical Witten effect explicitly by analyzing the fermion modes similarly to the original analysis of \cite{Callan:1982ah}.

\bigskip 
We would also like to comment on what our proposal implies for the massless limit. Due to the fact that the electric charge of the dyon degree of freedom is generated by the fermion vacuum, the charge radius is parametrically given by $R_c\sim 1/m_\psi$. In the massless limit, we see that the total charge of the dyon remains fixed, but that the charge radius goes to infinity and the charge density becomes infinitely dilute. In the strictly massless theory, this is similar to a DC shift of the vacuum electric charge, but this only strictly makes sense with an IR regulator. This is similar to the idea of \cite{Intriligator:2013lca} which suggests that the Hilbert space has different super-selection sectors corresponding to the plane-wave normalizable zero-energy modes so that fermion scattering shifts the vacuum state along the space of these super-selection sectors. In the case at hand, these super-selection sectors may come from to the $E=0$, $j=0$ fermion modes which are non-normalizable but generate a constant charge density.\footnote{
Since these modes contribute a finite charge density, they will generate infinite total charge in infinite volume which requires regulation/renormalization matching with our discussion above. }

\section{Monopole-Fermion S-Wave Scattering}
\label{sec:2D}

 Now we would like to study the scattering of spherically symmetric (i.e. $j=0$) fermion modes off of the spherically symmetric BPS monopole. 
 
 First, we would like to argue that there is a consistent truncation to the theory of $j=0$ modes. As we have discussed, all non-abelian modes of the UV gauge field except for the dyon degree of freedom are gapped out with an effective mass of order $m_W$ \cite{Brennan:2021ewu}.  This means that the $j=0$ fermion modes do not interact with the higher angular momentum modes of the $W$-boson in the monopole core. Since the dyon is a spherically symmetric mode, this implies that the interaction between the dyon degree of freedom and the fermions does not mix the fermionic angular momentum modes. Therefore an in-coming $j=0$ fermion mode can only scatter into an out-going $j=0$ fermion mode, while at most exciting the dyon due to conservation of angular momentum.\footnote{Note that the gauge field can mediate interactions between the different angular momentum sectors, but these are perturbatively small and we will ignore their effect in this paper.} Additionally, it also implies that the dyon can only decay into the $j=0$ modes. The reason is that we can think of the excited dyon state as the trivial in-coming state which has vanishing angular momentum and therefore the interaction only allows for radiation into the $j=0$ modes. 
This fact is also obvious since the source term is spherically symmetric and hence can only decay in a spherically symmetric way.\footnote{Here since we are interested in the monopole-fermion interaction, for now we will also ignore the fluctuations of the higher angular momentum modes of the dynamical $U(1)$ gauge field. We will comment on their effects in the case of general scattering in Section \ref{sec:GeneralScattering} where the interaction will simply be related to the effects of a propagating charged fermions  in $4d$. }
All together, this implies that there is a consistent truncation to the $j=0$ sector of the theory.

\bigskip 
Now, we would like to study in detail the scattering of a spherically symmetric fermion off of the monopole. 
This scenario will encapsulate much of the physics of and act as a warm up for the general fermion-monopole scattering which we will discuss in the following section. 

First we will derive the effective theory for massless fermions and then consider the effect of turning on a mass. Because of our previous discussion, the s-wave scattering can be studied in the truncated theory where we restrict to only the $j=0$ modes of the fermions. 
As first discussed by \cite{Callan:1982au,Callan:1982ah,Callan:1982ac}, the truncated theory is equivalent to an effective $2d$ theory of radially propagating fermion shells. 

To see this, first note that the $j=0$ angular momentum modes of the massless fermions can be decomposed as 
\eq{
\psi^A=\frac{U(\theta,\phi)\,e^{ -i \phi/2}}{r}\int dk\left(\begin{array}{c}
0\\e^{ik(t+r)}\chi_k^A 
\end{array}\right)~, \quad \tilde\psi^A=\frac{U(\theta,\phi)\,e^{i \phi/2}}{r}\int dk\left(\begin{array}{c}
e^{i k(t-r)}\tilde\chi_{k}^A\\0
\end{array}\right)~,
}
where $A=1,...,N_f$ and  
$\chi_k^A,\tilde\chi_k^A$ are the Grassman-valued mode operators and $U(\theta,\phi)$ is the frame rotation as described in the previous section in \eqref{framerotation}. 

Because of this decomposition, we can reduce the $4d$ fermions to effective $2d$ left- and right-moving fermions $\chi^A,\tilde\chi^A$:
\eq{
\psi^A=\frac{U(\theta,\phi)\,e^{-i \phi/2}}{r}\left(\begin{array}{c}
0\\
\chi^A(t+r)
\end{array}\right)\quad, \quad \tilde\psi^A=\frac{U(\theta,\phi)\,e^{i \phi/2}}{r}\left(\begin{array}{c}
\tilde\chi^A(t-r)\\0
\end{array}\right)~, 
} 
as described in \cite{Callan:1982au,Callan:1982ah,Callan:1982ac}. 
Plugging the $j=0$ mode expansion into the $4d$ UV Lagrangian leads to effective $2d$ theory of fermions propagating on the $(r,t)$ half-plane that interact with the dyon degree of freedom on the boundary at $r=0$ as \cite{Brennan:2021ewu}
\eq{\label{OGint}
S_{int}&=\int d^2x\, \delta(x)e^{-i \varphi}\bar{\tilde\chi}_A\chi^A+c.c.~. 
}
This boundary interaction imposes the boundary condition on the fermions
\eq{\label{OGBC}
\chi^A=e^{i \varphi}\tilde\chi^A\big{|}_{r=0}~. 
}
Note that the boundary condition comes from the non-abelian interaction with the $W$-boson and locally breaks $SU(N_f)^{(0)}\times SU(N_f)^{(0)}\to SU(N_f)^{(0)}$ and the $\IZ_{N_f}^{(0,1)}$ (0,1)-form global symmetry. This local ``breaking'' of the (0,1)-form symmetry is the statement that the dyonic lines are charged under the 1-form global symmetry equivalent to how charged local operators transform non-trivially under gauge transformations of a 0-form symmetry.

Now by doing a field redefinition 
\eq{\label{fieldredef}
\chi\to e^{ i \varphi/2}\chi\quad, \quad \tilde\chi\to e^{-i \varphi/2}\tilde\chi~, 
}
we can exchange this boundary condition/interaction for a bulk interaction term as in \cite{Polchinski:1984uw}:\footnote{Here we drop the Heaviside step function which is used as a regulator. Additionally, note that the spherical reduction of the gauge kinetic term in the above action comes with a factor of $r^2$ due to measure for spherical coordinates. The spatially dependent factor prevents the gauge theory from confining as is standard in $2d$ QED.  
}
\eq{
S=&\int d^2x\left( i \bar\chi\left(D_++\ihalf \dot\varphi \right)\chi+i \bar{\tilde\chi}\left(D_--\ihalf \dot\varphi \right)\tilde\chi  
+\frac{r^2}{2g^2}F_{tr}^2\right)\\&
+\frac{1}{2m_W} \int dt(\dot\varphi-2a_0)^2 ~.
}
 After the field redefinition \eqref{fieldredef}, the boundary conditions for the fermion fields are given 
\eq{\label{NewBC}
\chi=\tilde\chi\Big{|}_{r=0}~, 
}
which is still gauge invariant due to the fact that the field redefinition involved the Goldstone field $\varphi$.

\bigskip
Now we can consider the effect of turning on a mass term. In the spherical reduction, this becomes a $2d$ mass term
\eq{
\CL_{4d}=m_\psi \psi^A\tilde\psi_A+c.c.\quad \longrightarrow \quad \CL_{2d}=m_\psi \chi^A\tilde\chi_A+c.c.~.
}
However, in rewriting the mass term in a $SU(N_f)^{(0)}$ symmetric way, we also had to redefine our fields \eqref{psisplitting} so that the gauge interaction now becomes  
\eq{\label{OGint}
S_{int}&=\int d^2x\, \delta(x)e^{-i \varphi}\epsilon_{AB}\bar{\tilde\chi}^A\chi^B+c.c.~.
}
where here $\epsilon_{AB}=\mathds{1}_{n_f}\otimes \sigma^1$. 
This interaction leads to the boundary condition on the fermions
\eq{\label{OGBC}
\chi^A=\epsilon^{AB} e^{i \varphi}\tilde\chi_B~. 
}
Again, this boundary condition breaks the 0-form global symmetry $SU(N_f)^{(0)}\to SU(n_f)^{(0)}$ to its UV form and breaks $\IZ_{N_f}^{(0,1)}$ (0,1)-form global symmetry and implies that the dyonic lines are charged under the (0,1)-form global symmetry.

\bigskip 
It will also be instructive to study the bosonized theory. Here we will bosonize the fermions as\footnote{As stressed in \cite{vanBeest:2023dbu}, here we are simply bosonizing the Cartan of the full $U(N_f)$ current algebra generated by the fermions. 
}
\eq{\label{fermbc}
\bar\chi_A\chi^A=D_+ H_A\quad, \quad \bar{\tilde\chi}^{A}\tilde\chi_A=D_-\tildeH_A~,
}
for fixed $A$ and $H_A,\tildeH_A$ are left-/right-moving chiral bosons. The boundary conditions \eqref{fermbc} are given in terms of the $H_A,\tildeH_A$
\eq{\label{Hbcmassless}
H_A=\tildeH_A\big{|}_{r=0}~,
}
for the massless case and 
\eq{\label{Hbcmass}
H_A=\epsilon_{AB}\tildeH_B\big{|}_{r=0}~,
}
for the massive case. As in the fermionic description, these boundary conditions break $SU(N_f)^{(0)}\times SU(N_f)^{(0)} \to SU(N_f)^{(0)}$ and $SU(N_f)^{(0)}\to SU(n_f)^{(0)}$ respectively but preserve gauge invariance due to the field redefinition \eqref{fieldredef}. 

We can then write the action in terms of the complex scalar fields $h_A=H_A+\tildeH_A$ as 
\eq{\label{2dbosaction1} 
S=\int d^2x\left(|\partial h|^2-F_{tr}  h+\frac{r^2}{2g^2}F_{tr}^2\right)+\int dt\left(\frac{(\dot\varphi-2a_0)^2}{2m_W}+\frac{(\dot\varphi-2a_0)}{4\pi}h 
\right)
} 
for the massless case and 
and as 
\eq{\label{2dbosaction2} 
S=&\int d^2x\left(|\partial h|^2-F_{tr} h-m_\psi^2\cos\left(\frac{h_A}{2}\right)+\frac{r^2}{2g^2}F_{tr}^2\right)\\&
+\int dt\left(\frac{(\dot\varphi-2a_0)^2}{2m_W}+\frac{(\dot\varphi-2a_0)}{4\pi}h 
\right)
}
for the massive case where $h=\sum_A h_A$. It will also be useful to consider the decomposition of $h(t,r)$ into left- and right-handed components: $h=\Theta(t-r)+\tilde\Theta(t+r)$.  

As discussed in \cite{vanBeest:2023dbu,Callan:1982au}, there are quantum corrections to the effective action. The corrections investigated there can be interpreted as the 
1-loop contribution to the gauge-mediated 4-fermion interaction restricted to the $j=0$ fermion modes.\footnote{Note that there is no 2-fermion mass term generated for massless fermions at 1-loop order by gauge interactions due to symmetry.} In the $2d$ effective action, this becomes a 4-fermion term which, upon bosonization, maps to a cosine-potential with coupling strength that goes like $g^2/r^2$. However, this 1-loop correction couples the $j=0$ modes to all higher angular momentum modes at the same order in perturbation theory. This interaction requires further analysis as we do not usually expect 4-fermion interactions to gap out fermions in $4d$. Here, we will consider the weak coupling limit and ignore these corrections, leaving a more complete analysis for future investigation.\footnote{Another reason why we may be able to safely ignore these corrections is the following. As discussed in \cite{vanBeest:2023dbu,Callan:1982au}, one of the corrections lead to a repulsive $1/r^2$ potential. In $2d$, this is a short range potential -- it corresponds to a $1/r^4$ potential in $4d$. We expect that the physics with a short-range repulsive potential is similar to the interacting resonant model (IRM) in $2d$ \cite{RLM1,RLM2}. The IRM theory describes $2d$ fermions interacting with a defect with a delta function repulsive potential. The IRM model is integrable and has been solved exactly -- the bulk fermion modes can tunnel through the short-range repulsive potential to interact with the defect degrees of freedom. We expect that a similar physical picture may occur here due to the short range nature of $1/r^2$ potentials in $2d$. }

Now we would like to consider the $\IZ_\Nf^{(0,1)}$ (0,1)-form global symmetry. As we discussed, the $\IZ_\Nf^{(0,1)}$ global symmetry acts on $\varphi$ since $\dot\varphi\mapsto \dot\varphi+2\Lambda_1$ under large gauge transformations. 
In the field redefinition \eqref{fieldredef} we shift the bulk fields by the dyon degree of freedom so that the fermions, which originally are invariant under the diagonal $\IZ_\Nf$ subgroup of $U(1)_{\rm gauge}$ and $SU(N_f)\times SU(N_f)$ or $SU(N_f)$, now transform non-trivially under $\IZ_\Nf^{(0,1)}$ transformations
\eqref{flatzn}:
\eq{
dH_A\longmapsto dH_A+\Lambda_1~, \quad d\tildeH_A\longmapsto d\tildeH_A-\Lambda_1~, \quad \dot\varphi\longmapsto \dot\varphi+2\Lambda_1~,\quad a_0\longmapsto a_0+\Lambda_1~.
}
To determine the intrinsic action on the scalar fields, one must realize that in a non-trivial $\IZ_\Nf^{(0,1)}$ background, $H_A,\tildeH_A,\varphi$ are sections of a $U(1)$-bundle. We can think of them as defined in local patches where the $\IZ_\Nf^{(0,1)}$ background gives a rule for transformation between the patches by a $\IZ_\Nf^{(0)}$ action:
\eq{\label{flatzn}
H_A\longmapsto H_A+\frac{2\pi n}{N_f}\quad, \quad \tildeH_A\longmapsto \tildeH_A-\frac{2\pi\tilde{n}}{N_f}\quad, \quad \varphi\longmapsto \varphi+\frac{4\pi}{N_f}~. 
}
Importantly, the fields $\Theta,\tilde\Theta$ are invariant and hence define global sections because they have charge $N_f$ under the $\IZ_\Nf$ symmetry.

\subsection{Dyon Decay}

Now we will look more carefully at the process of dyon decay. We will now explicitly demonstrate that the conserved (0,1)-form $\IZ_\Nf^{(0,1)}$ global symmetry implies that the dyon does not decay unless it has vanishing charge  mod$_{N_f}$. 

Let us again consider the quantum mechanics describing the dyon degree of freedom. This is essentially given by the charged particle on the ring coupled to an external magnetic flux:
\eq{
L=\frac{1}{2m_W}(\dot\varphi-2a_0)^2+\frac{\theta}{2\pi} (\dot\varphi-2a_0)~,
}
where $\theta$ is a topological theta angle. As we have discussed in Section \ref{sec:Monopoles}, this theory has a conserved momentum $\hat{p}=\frac{1}{m_W}(\dot\varphi-2a_0)+\frac{\theta}{2\pi}$ which grades the Hilbert space:
\eq{
\CH=\span\Big\{|n\rangle~,~n \in \IZ~\Big{|}~\hat{p}|n\rangle=n|n\rangle\Big\}~.
}
Here the Hamiltonian is given by 
\eq{
\hat{H}=\frac{m_W}{2}\left(\hat{p}-\frac{\theta}{2\pi}\right)^2+2a_0\hat{p}~.
}
We can then see that each state $|n\rangle$ has an energy
\eq{
E_n=\frac{m_W}{2}\left(n-\frac{\theta}{2\pi}\right)^2~,
}
and sources a bulk $A_0$ corresponding to a point particle of charge $n$ which is consequently charged under the $\IZ_N^{(0,1)}$ symmetry due to the transformation properties of $A_0$. We can see directly that the state $|n\rangle$ has charge $2n$ under a $\IZ_N^{(0,1)}$ as follows.  Due to the fact that the last term in the Hamiltonian transforms under $\IZ_N^{(0,1)}$ when acting on the state $|n\rangle$, when we integrate out the quantum mechanical degree of freedom in the state $|n\rangle$, the contribution to the partition function will transform as a Wilson line of charge $2n$ which transforms under the $\IZ_N^{(0,1)}$ symmetry.

In the monopole-fermion scattering, $\varphi(t)$ couples to a dynamical $\theta$ which is the boundary value of a $2d$ periodic field 
\eq{
\theta=\frac{h}{2}\big{|}_{r=0}=\half\left(\Theta+\tilde\Theta\right)\big{|}_{r=0}~.
} 
Note here that $h,\Theta,\tilde\Theta$ are all invariant under the $SU(N_f)^{(0)}\times SU(N_f)^{(0)}$ or $SU(N_f)^{(0)}$ flavor symmetries as appropriate and that $\Theta,\tilde\Theta$ have gauge charges $\pm N_f$ respectively.

The coupling between  $\varphi(t)$ and $\Theta+\tilde\Theta$ leads to an instability of the quantum mechanical theory. This instability allows the energy levels in \eqref{QMEnring} to decay to $|0\rangle$ by  emitting bulk solitons. For example, the boundary value of $\Theta+\tilde\Theta$ can jump by $\pi N_f$ by emitting a $\tilde\Theta$ (winding) kink solution. 
This shifts the energy levels by $n_f$ units as can be seen from the energy equation \eqref{QMEnring}.  

One may then be tempted 
to conclude that the dyon can decay by a single energy level by a kink where $\tilde\Theta\to \tilde\Theta+4\pi$. However, such a process will violate the global symmetries of the theory. In the following, we will describe the dynamics in terms of chiral fields. Let us parametrize the scalar degrees of freedom in terms of $\Theta,\tilde\Theta$ as above and:
\footnote{Here there is a $\IZ_{N_f}$ redundancy from the fact that 
\eq{
\Theta=\sum_n n \,\Theta_n~{\rm mod}_{N_f}\quad, \quad \tilde\Theta=\sum_n n\, \tilde\Theta_n~{\rm mod}_{N_f}~.
}
This  $\IZ_{N_f}$ relation is enforced by a $\IZ_{N_f}$ gauge field under which both sides of the above equation have charge 1. Note that there is not a unique charge assignment of the $\Theta_i,\tilde\Theta_i$ due to the individual periodicity of the $\Theta_i$. The choice is equivalent to choosing a representative of the generator of the center of $SU(N_f)^{(0)}$. One can make the simple choice that $\Theta_1,\tilde\Theta_1$ have charge 1 and all other $\Theta_i,\tilde\Theta_i$ are uncharged. }
\eq{ 
\Theta_i=H_i-H_{i+1}~, \quad \tilde\Theta_i=\tildeH_i-\tildeH_{i+1}~, \quad i=1,...,N_f-1~.
}
The periodicity of the new fields is given by 
\eq{\label{thetabc}
\Theta\sim \Theta+2\pi N_f\quad, \quad \Theta_i \sim \Theta_i+2\pi N_f~,
}
and similarly for $\Theta,\tilde\Theta_i$ with boundary conditions:
\eq{
\Theta-\tilde\Theta\big{|}_{r=0}=0\quad, \quad \Theta_A-\tilde\Theta_A\big{|}_{r=0}=0~.
}

\bigskip Now let us consider the charge 2 dyon state. This state is uncharged under the flavor $SU(N_f)^{(0)}$ or $SU(N_f)^{(0)}\times SU(N_f)^{(0)}$ global symmetry but, as we have discussed, carries charge under the $\IZ_{N_f}^{(0,1)}$ (0,1)-form global symmetry. The dyon degree of freedom couples to the bulk fields $\Theta,\tilde\Theta$ as well as possibly indirectly to 
$\Theta_i,\tilde\Theta_i$. The fields $\Theta_i,\tilde\Theta_i$ are all charged under the global 0-form symmetry $SU(N_f)^{(0)}$ or $SU(N_f)^{(0)}\times SU(N_f)^{(0)}$ as appropriate while the fields $\Theta,\tilde\Theta$ are uncharged. Additionally, following our previous discussion we also see that all of the bulk fields $\Theta_i,\tilde\Theta_i,\Theta,\tilde\Theta$ are uncharged under the $\IZ_{N_f}^{(0,1)}$ global symmetry.  
 This means that any dyon state which has non-trivial charge under the $\IZ_\Nf^{(0,1)}$ global symmetry cannot decay to any of the bulk $\Theta,\tilde\Theta$ modes.  
Together, the conservation of the 0-form flavor symmetry and the (0,1)-form  symmetry forbids the decay of the dyon level except by multiples of $N_f$ via a $\tilde\Theta$-winding mode. 

Another way to rephrase the fact that there is a preserved $\IZ_\Nf^{(0,1)}$ global symmetry, is the statement that the fermion-dyon interaction term does not facilitate the radiation of bulk fermion modes with charges less than $N_f$ due to the preserved $SU(N_f)^{(0)}\times SU(N_f)^{(0)}$ or $SU(N_f)^{(0)}$ symmetry. The reason is that fermion modes can only carry integer charge, so there is no way to divide up $Q\neq k N_f$ electric charge among $N_f$ fermions in an $SU(N_f)^{(0)}\times SU(N_f)^{(0)}$ or $SU(N_f)^{(0)}$ symmetric way.

As described in \cite{vanBeest:2023dbu}, the out-going state described in the standard Callan Rubakov effect corresponds to the $2\pi$ $\tilde\Theta$-kink which is generated by $e^{ i \frac{\tilde\Theta}{N_f}}$. However, this operator is not well defined since it is not invariant under $\tildeH_A\sim \tildeH_A+2\pi$ which is a type of gauge symmetry. The proposal of \cite{vanBeest:2023dbu} is that this issue can be cured by attaching a line operator which compensates for the gauge non-invariance. Physically, this is equivalent to taking the $N_f^{th}$-root of a well defined operator $e^{i \tilde\Theta}=\prod_A\tilde\chi_A$ which introduces a co-dimension 1 space time branch cut in all correlation functions.

\subsection{Scattering and Energetics}

Now let us describe what happens when we scatter a single fermion mode off of the monopole.  
Consider sending in a $j=0$ fermion state for $\psi^A$ with fixed $A$. In the bosonized description, this is the same thing as sending in a kink configuration for $H_A$ that winds by $2\pi$. This can be seen from the action \eqref{2dbosaction1} -- \eqref{2dbosaction2} where the coupling between $h_A=H_A+\tildeH_A$ and the gauge field is equivalent to the coupling of the gauge field to the winding current of $h_A$. 

Now from the boundary conditions on $H_A$ \eqref{Hbcmassless} -- \eqref{Hbcmass}, we see that the incoming kink reflects as a $\tildeH_A$ kink. This shifts the effective $1d$ $\theta$-angle by $2\pi$ and shifts the boundary state momentum by 1 unit. 
This means that we have effectively excited the dyon degree of freedom to the $n=1$ state so that the monopole has now become a dyon of charge $2$:
\eq{
\psi^A+M~\longrightarrow \tilde\psi^A +D~.
}
Now we would like to explain the most important feature of this scattering: how does an incident fermion impart gauge charge onto the monopole? First, we will consider the case of massive fermions -- we will then discuss the massless limit $m_\psi\to 0$. 

Let us address the fact that the monopole-fermion scattering can impart a $\IZ_\Nf^{(0,1)}$ global charge on the monopole. This fact may appear strange because the fermion fields are invariant under the $\IZ_\Nf^{(0,1)}$ symmetry. However, the crucial point is that physical scattering states are necessarily gauge invariant. These are constructed by dressing the bare (i.e. gauge non-invariant) fermion scattering state by gauge fields which allows them to transform non-trivially under the $\IZ_\Nf^{(0,1)}$ symmetry. These states can be thought of as being created by fermion fields that are connected to Wilson lines and hence can transform under the (0,1)-form  symmetry because the operator that creates the state is now no longer local. Said differently, asymptotic scattering states are gauge invariant, but carry $SU(N_f)^{(0)}$ (or $SU(N_f)^{(0)}\times SU(\Nf)^{(0)}$) charge and therefore must transform non-trivially under $\IZ_\Nf\subset SU(\Nf)^{(0)}$ (or $\IZ_\Nf\subset SU(N_f)^{(0)}\times SU(N_f)^{(0)}$). This means that the asymptotic scattering states can carry and impart $\IZ_{N_f}^{(0,1)}$ (0,1)-form global charge.

Next we would like to discuss in detail how gauge charge is imparted to the monopole at parametrically low energies. As we have mentioned the coupling between the effective $2d$ fermions and the dyon degree of freedom can be written as a dynamical $1d$ $\theta$-angle,  the interaction can induce a shift of the energy levels of the QM by a $1d$ Witten Effect. As we have discussed, an in-coming fermion mode is described by an $H_A$-kink which winds by $2\pi$ and the out-coming mode is a $\tildeH_A$-kink which winds by $2\pi$ so that the effective $1d$ $\theta$-angle in the dyon quantum mechanics shifts by $2\pi$. This leads to a shift in the energy levels so that the quantum mechanics ground state shifts from $|0\rangle\to |1\rangle$ and turns the monopole into a dyon of charge 2.

In terms of the bulk $4d$ fields, the   $H_A,\tildeH_A$ fields describe the phase mode of the $\psi^A,\tilde\psi^A$ $j=0$ modes. Thus, a bosonic kink describes a propagating phase shift  of $\psi^A,\tilde\psi^A$ by $2\pi$ which induces a $4d$ $\theta$-angle: $\theta_{4d}=4\pi$. As we discussed in Section \ref{sec:symmetries}, this induces an electric charge on the monopole by a shift in the fermion vacuum charge which is distributed in a cloud around the monopole of radius $R_c\sim 1/m_\psi$. This bulk electric field contributes to the mass of the dyon 
\eq{\label{dyonenergy}
\Delta E_{\rm dyon}\sim m_\psi~,
}
so that exciting the dyon degree of freedom/fermion vacuum via the shift in $\theta$ only requires energy scales that are comparable to the energy required for fermion scattering. 
 
After scattering, the dyon charge sources an electric field that interacts with the out-going fermion modes. Since the charge cloud and out-going fermions have opposite charge so that the dyon can trap the out-going fermions in its electrotatic potential. As discussed in \cite{Zhang:1988ab,Ravendranadhan:1989vp,Tang:1982fc}, the dyon has bound states that are similar to that of the hydrogen atom with binding energy that  scales parametrically as 
\eq{
E_{\rm binding}\sim m_\psi~.
}
This suggests that incident fermions will be captured by the excited dyon if they are scattered at energy scales of $E_{\rm scatt}\lesssim 2m_\psi$ whereas for $E_{\rm scat}>>m_\psi$, fermions will have enough energy to escape the dyon's electrostatic potential.

 \bigskip
 In the case of massless fermions where $m_\psi\to 0$, the dyon can be interpreted as a monopole with an infinitely disperse cloud of fermions (carrying electric charge). Additionally, in this limit the binding energy of the dyon goes to zero which means that any scattered fermion will impart a charge by shifting the fermion vacuum and then will escape off to infinity.  It is possible that this may be realized in the renormalization of the infinite vacuum contribution to the charge density.  
 
{This suggests that we can think of the Hilbert space of our theory as belonging to a family of Hilbert spaces that are indexed by this vacuum electric charge}. Such a family of Hilbert spaces with varying vacuum charge density naturally arises in our theory from considering the family of vacuum states that arise from quantizing the zero energy mode of the $j=0$ fermion modes.\footnote{Note that the zero energy mode belongs to a continuum of scattering states and so it may seem strange to separate this mode from the others in the continuum which can be used to construct smooth states by Fourier transform smooth distributions. However, since the zero energy  mode  is at the bottom of the continuum, all smooth states will be constructed by the Fourier transform of a function $F(k)$ with $\lim_{k\to 0}F^{(n)}(k)=0$ $\forall n\geq0$ and hence in a sense is distinct from the continuum scattering states. }
 As discussed in \cite{Intriligator:2013lca}, these modes are not usually quantized because they are non-normalizable and hence lead to disconnected super-selection sectors.  
We believe these subtle issues are the intrinsic source of confusion about the Callan Rubakov effect and deserve further investigation.

\bigskip 
Now we would like to discuss the  decay of the dyon degree of freedom. 
Recall that the $\IZ_{N_f}^{(0,1)}$ (0,1)-form symmetry only protects the dyon from decaying mod$_{N_f}$which allows for the emission of electric charge in multiples of $N_f$. And indeed, we expect that this is energetically favorable due to the fact that the dyon can decay through angular momentum modes with energies that are arbitrarily close to $m_\psi$. This should also be expected from the fact that integrating out the dyon degree of freedom gives a local `t Hooft operator \cite{Fan:2021ntg}:
\eq{
V_{\rm `t~Hooft}\sim e^{i \frac{\Theta+\tilde\Theta}{2}}\delta(r)~,
}
which has the effect of transforming an in-going state with $n_f$ fermions to an out-going state with $n_f$ fermions with the opposite chirality. 

From our previous analysis, we see that the dyon quantum mechanics can only emit gauge charge via $\tilde\Theta=\sum_A \tildeH_A$ winding modes which are the combined winding mode of each $\tildeH_A$. Upon returning to the fermionic description, this is the statement that the dyon will decay by sending out a single quanta of each $\tilde\psi_A$. 

In the original setting of the Callan Rubakov effect, $SU(5)$ GUT theory with a single generation, the interaction of Standard Model fermions with the  minimal spherically symmetric monopole (with respect to $U(1)_Y$ hypercharge) reduces to the interaction of $N_f=4$ Weyl doublets interacting with the spherically symmetric $SU(2)$ monopole.\footnote{Here we pick the magnetic charge to be embedded as $T_3={\rm diag}(0,0,1,-1,0)$ as in \cite{Callan:1983tm,Callan:1982au,Callan:1982ac,Gardner:1984zd,Gardner:1983uu,Brennan:2021ewu}.} Here there is an effective $SU(4)\times SU(4)$ flavor symmetry in the effective $SU(2)\to U(1)_Y$ gauge theory which plays the role of the global 0-form symmetry in our (and the standard Callan Rubakov) analysis.\footnote{Note that the $SU(3)\times SU(2)$ gauge symmetry does not embed into the $SU(4)\times SU(4)\times U(1)_{\rm gauge}$ due to the fact that the magnetic charge does not commute with the generators of $SU(3)\times SU(2)$. }
  Our analysis shows that the process 
\eq{
2\bar{p}+M~\longrightarrow~ 3e^-+p+M~,
}
does indeed occur, but that scattering a single proton may be captured by the monopole:
\eq{
\bar{p}+M~\longrightarrow~ (e^-+2u+D)~,
}
depending on energetics, 
where $D$ is the dyon and $(\,\underline{~~}+D)$ indicates a fermion-dyon bound state. 

Now we would like to make a few comments about the application of our analysis to scattering processes in the Standard Model: 
\begin{itemize} 
 \item Here our analysis only strictly applies {at energies $E_{EWSB}\gg E\gg\Lambda_{QCD}$ and }when the fermions all have identical mass. This would require that the Yukawa couplings be diagonal and identical, and that the CKM matrix is trivial. It would be interesting to study how our proposal is modified by weak symmetry breaking effects.  

\item As discussed in the introduction, both of these processes lead to baryon and lepton number violation. In the second process, the dyon carries $U(1)_{B-L}$ charge due to the fact that the generator of the $U(1)_{B-L}$ in the $SU(5)$ theory is given by 
\eq{
Q_{B-L}=Q_Y+Q_D
}
where $Q_Y$ is the generator of hypercharge and $Q_D$ acts on the $SU(5)$ fermion multiplets with charges $Q_D[\mathbf{\bar{5}}]=-3$ and $Q_D[\mathbf{10}]=1$. Since the monopole does not carry charge with respect to $Q_D$, but does with respect to $Q_Y$, the dyon can carry charge with respect to $Q_{B-L}$. When enough charge is accumulated, the $U(1)_{B-L}$ charge is radiated away, and thus $U(1)_{B-L}$ is conserved as in the original Callan Rubakov effect.

\item  Here, it may seem strange that the down quarks carry color charge whereas the anti-proton carries no color charge. The reason is that actually the spherically symmetric monopole is also embedded non-trivially along the $SU(3)$ part of the Standard Model gauge group and thus in a sense carries color charge itself.

\item Note that as discussed in \cite{Aharony:2022ntz}, Wilson line operators of large electric charge $Q\sim \frac{1}{g^2}$ are screened in QED. Here we expect a similar behavior and thus, there may be additional effects for the Callan Rubakov effect when $N_f\gtrsim \frac{1}{g^2}$. For this reason, we only claim our results hold for $N_f\ll \frac{1}{g^2}$. 
\end{itemize}

\noindent In summary, we see that our analysis produces the original Callan Rubakov effect for s-wave scattering for asymptotic states with integral fermion number (i.e. that monopoles can catalyze proton decay/baryon number violation) but modifies it so as to avoid the fractional fermion states.

\section{General Monopole-Fermion Scattering}
\label{sec:GeneralScattering}

It is now somewhat straight forward to analyze the scattering for general fermions  beyond the $j=0$ mode. Due to the spherical symmetry of the monopole, all higher angular momentum modes can be reduced to effective $2d$ fermions on the half space. 
To write down the effective theory, we simply need to understand the boundary conditions for the higher angular momentum modes. 
This is dictated by the tree level interaction of the UV theory of the angular momentum modes with the $W$-boson. As we have discussed, the fact that all higher angular momentum modes of the non-abelian degrees of freedom are gapped out with a mass of order $m_W$ implies that  
the dyon degree of freedom is only allowed to decay into the $j=0$ modes as discussed in the previous section. 

The effective interaction is then determined  by 
 \eq{\label{TreelevelW}
 S_{int}=\int e^{i \varphi}\,\bar\psi\, T_+ (\gamma^{\theta}+i \gamma^\phi)\psi\,M(r)\, d^4x+c.c. 
 }
 Again, since the dyon degree of freedom is spherically symmetric, this interaction does not mix the angular momentum states. Therefore, an asymptotically in-going mode with angular momentum $|j,m\rangle$ will scatter into an out-going state of the same $|j,m\rangle$. 
 
The tree level interaction between the fermions and dyon degree of freedom is determined explicitly by the mode expansion as in \eqref{asymptoticmassivetotal1} -- \eqref{asymptoticmassivetotal2},   
which have $j$-angular momentum spherical wave modes that propagate both in- and out-ward.  
Using the explicit expressions in Section \ref{sec:Monopoles} we see that in the monopole background, the independent, orthogonal modes generate an asymptotic current flux at infinity as in \eqref{TotalModeFlux}: 
\eq{
\Delta Q=\int_{S^2_\infty}\bar\Upsilon \gamma^\mu \Upsilon\, \hat{r}_\mu\, d^2\Omega=2(a_1^\dagger a_1+a_2^\dagger a_2-a_3^\dagger a_3-a_4^\dagger a_4)~.
}
This above analysis captures the tree level interaction of the fermions with the dyon degree of freedom at tree level. 
Thus, we find that if we send in a quanta of $a_1,a_2$ ($a_3,a_4$), then we deposit $+2$ ($-2$) electric charge and excite the dyon. Note that although the dyon degree of freedom couples to IR modes which have positive and negative asymptotic current (unlike the $j=0$ modes), it cannot radiate away its charge through these modes. The reason is that such a decay is prevented by conservation of angular momentum -- this forces the dyon to only decay through the $j=0$ modes.

Our analysis for the $j=0$ modes suggests that this deposits electric charge by again inducing a Witten effect on the monopole. 
Here, the crucial observation is that the phase mode of the fermions carries their gauge charge. Separating out the phase mode from the other parts of the fermion generates an axion field according to the transformation properties of the original fermion field. This means that scattering a charged fermion -- i.e. sending in gauge current -- contains a winding mode of this effective axion field which then induces a    shift in the $\theta$-angle, thereby depositing electric charge in a cloud of radius $R_c\sim 1/m_\psi$.\footnote{One could in principle truncate to the $j$- and 0- angular momentum modes which is described by a consistent effective $2d$ theory. Due to the universality of the coupling between $\varphi$ and the $2d$ fermions \eqref{TreelevelW}, the analysis should proceed almost identically to that of Section \ref{sec:2D} where the incoming $j\neq 0$ bosonic kinks induces a shift in the effective $\theta_{1d}$, thereby converting the monopole into a dyon. Again, these bosonic kinks describe the phase mode -- i.e. axionic mode -- of the bulk $4d$ fermions which leads to the conservation of electric charge in $4d$ via the dynamical Witten effect. 
}

Again, the dyon can then either decay or not according to the conserved $\IZ_\Nf^{(0,1)}$ (0,1)-form global symmetry. As discussed in detail in the previous section, the dyon can decay into the fermionic $j=0$ modes due to spherical symmetry.

This gives us the general picture of scattering a fermion off of a monopole. Here, the fermions will excite the dyon via the tree-level UV interaction with the (spherically symmetric) trapped $W$-boson. This can then decay (or not) to the spherically symmetric modes fermion modes:
\eq{
\psi+M~&\longrightarrow~(\tilde\psi+D)~,\\
\psi^{n_f}+M~&\longrightarrow~\tilde\psi^{n_f}+\prod_A\bar{\tilde\psi}^A +M~,
}
where the asymptotic out-going angular momentum modes for the second process are given in terms of the normalizable states at $r\to 0$ as in \eqref{asymptoticmassivetotal1}.

This implies the general physical picture for monopole-fermion scattering. When any fermion wave packet is sent into the monopole core, it will interact with the dyon degree of freedom via the tree level UV coupling of the fermion to the $W$-boson. Here the coupling to the $4d$ photon (and its higher angular momentum modes) plays no special role -- we simply have a propagating charged fermion in QED. 
The incident fermion will scatter as the UV charge conjugate field with respect to the UV $SU(2)$ gauge symmetry. The scattering will also impart a charge onto the monopole via a shift in the fermion vacuum which leads to a charge cloud of radius $R_c\sim 1/m_\psi$. This charge will enforce charge conservation and may or may not trap the out-going fermion due to the generated electrostatic potential.

\bigskip 
Again, in the massless limit, this picture becomes much more subtle. As before, the electric charge of the dyon is concentrated in a fermion cloud of radius $R_c\sim 1/m_\psi$ around the monopole core due to the Witten effect. As in the case of $j=0$ scattering, the limit $m_\psi\to 0$ sends the radius of this cloud to infinity and sends the local charge density to zero. Similarly, as discussed in \cite{Zhang:1988ab,Ravendranadhan:1989vp,Tang:1982fc}, the binding energy of the dyon-fermion system for finite angular momentum modes scales as $E_{\rm bind}\sim m_W$ so that again in the limit $m_\psi\to 0$ there is no possibility of a bound state and all scattered fermions escape to infinity. This leads to the same conclusion as in the scattering of $j=0$ fermion modes that scattering of charged fermions off of a monopole is dictated by the tree level interaction and that the conservation of gauge symmetry may  possibly be attributed to  shifting the super-selection sector of the Hilbert space, although this requires regularization in order to be made sensible.

\section{Discussion: Implications for Standard Model and Future Directions}

Here we have proposed a comprehensive picture of the monopole-fermion interaction {for massive fermions}. We have discussed  the interaction for massive particles at all angular momentum and what this implies for the interpretation of the massless monopole-fermion interaction. Our discussion suggests that the confusion surrounding the original analysis of the Callan Rubakov effect is associated to an issue with IR regulation { and that the Hilbert spaces are separated into super-selection sectors by an electric charge of the fermion vacuum}. 

We expect that the new physical effect of fermion trapping by monopoles should lead to a new class of phenomenological effects. Even though the Callan Rubakov effect still occurs -- i.e. that there are baryon number violating interactions which are catalyzed by monopoles -- we expect that this effect leads to less efficient baryon number violation in cosmological scenarios than originally proposed. The reason is that after a monopole participates in a baryon violating process, it acquires an electric charge which favors un-doing the process.  
This suggests that baryon number violating processes will be less efficient than originally proposed and hence may lead to a less efficient Langacker-Pi scenario which washes out early universe baryon number asymmetry \cite{Langacker:1980kd,Davis:1992ca}. 

Additionally, we expect interesting effects from the fact that the dyons generated by the monopole-fermion interaction can carry $U(1)_{B-L}$ charge. This means that interactions with the monopoles could generate a universe which appears to have a net $U(1)_{B-L}$ charge by constructing a scenario in which monopoles accrue $U(1)_{B-L}$ charge and then are inflated away.

Our results also suggest many interesting future directions. 
Besides generalizing to more complex spherically symmetric monopoles and higher dimensional fermion representations, the logic of this paper can be generalized (with some modifications) to describe the interaction of fermions with non-spherically symmetric monopoles. There, the interaction between the fermions and dyon degree of freedom will mix angular momentum modes according to the angular momentum of the monopole itself. Similarly, it should also have the ability to decay into different angular momentum modes. We believe that the dynamical Witten effect is the key to understanding the more general interaction. 

We also believe it would be interesting to further study the details our proposal such as the life time of the dyon after accumulating $N_f$ electric charge as well as the detailed structure of fermion bound- and scattering-states in the presence of the dyon.\footnote{We believe that there will be an interesting structure similar to that described in \cite{Iengo:2002eb}.} Additionally, it would be interesting to study the effects of 1-loop quantum corrections to the interaction and study the effects of small symmetry breaking effects. These would be especially interesting to the application of the Standard model due to the unequal masses in nature.

It is also interesting to ask how the Callan Rubakov effect is modified when the fermions are given a mass by a GUT scale Yukawa interaction. In this case, the monopole actually supports fermion zero-modes which may lead to a different physical effect. Interestingly, we do not expect that this effect occurs in supersymmetric Yang-Mills (SYM) theories without bare masses. The reason is that in these theories, the effective mass of the fermions is given by $m_\psi=m_W$ so that any fermion scattering occurs at high energies outside of the $U(1)$ effective field theory. However, this effect may be visible in SYM theories with bare masses $m_0$. In these theories, the effective fermion masses are shifted $m_\psi=m_W\pm m_0$ so that by taking large $m_0$, we can generate low effective mass fermions. 

It would also be interesting to better understand the role that the higher group and more general categorical symmetries play in the case of massless fermions. Although we have proposed a physical interpretation for the scattering process for massless fermions, it would still be interesting to better understand the subtleties associated with this limit. 

Another interesting application of this work is to the study of black hole physics. Recently, \cite{Maldacena:2020skw} showed how magnetically black holes can also lead to baryon number violation in the Standard Model due to electroweak symmetry restoration near the black hole horizon. Additionally, in \cite{Alford:1992ef}, the authors showed that there is a Callan Rubakov-type enhancement of s-wave scattering of fermions off of magnetically charged black holes. In light of our proposal, we believe it is would be interesting to revisit these ideas. 

\section*{Acknowledgements}
The author would like to thank Seth Koren, Sungwoo Hong, Konstantinos Roumpedakis, Pierluigi Niro, Andrea Antinucci, Giovanni Galati, Giovanni Rizi, Ken Intriligator, Zohar Komargodski, Marieke Van Beest, Diego Delmastro, Matt Reece, Liantao Wang, Colin Rylands, Gautam Satishchandran, David Tong, 
John Terning, Jeffrey Harvey,  Aneesh Manohar, Jorge Russo, Clay Cordova, and Thomas Dumitrescu for helpful discussions and related collaborations. TDB would also especially like to thank Seth Koren and Aneesh Manohar for comments on the draft. 
TDB is supported by Simons Foundation award 568420 (Simons
Investigator) and award 888994 (The Simons Collaboration on Global Categorical Symmetries).

\bibliographystyle{utphys}
\bibliography{TrappedCallanRubakovBib}

\end{document}